\documentclass[entropy,article,accept,pdftex,moreauthors]{Definitions/mdpi} 

\firstpage{1} 
\pubvolume{27}
\issuenum{4}
\articlenumber{385}
\pubyear{2025}
\copyrightyear{2025}
\externaleditor{Michael J. Way} 
\datereceived{10 March 2025} 
\daterevised{29 March 2025} 
\dateaccepted{1 April 2025} 
\datepublished{4 April 2025} 
\hreflink{https://doi.org/\linebreak 10.3390/e27040385} 
\pdfoutput=1 

\newcommand{\djs}{\mathcal{D}_{\rm{JS}}}
\newcommand{\dkl}{\mathcal{D}_{\rm{KL}}}
\newcommand{\classoption}[1]{\texttt{#1}}
\usepackage{array} 
\usepackage{makecell} 

\Title{\textls[-20]{An Informational--Entropic Approach to Exoplanet Characterization}}

\TitleCitation{\textls[-25]{An Informational--Entropic Approach to Exoplanet Characterization}}

\Author{{}{Sara} 
 Vannah $^{1,}${}{*}
\orcidA{}, Ian D. Stiehl $^{2}$ and Marcelo Gleiser $^{2,3}$\orcidB{}}

\AuthorNames{Sara Vannah, Ian D. Stiehl and Marcelo Gleiser}

\isAPAStyle{%
       \AuthorCitation{Lastname, F., Lastname, F., \& Lastname, F.}
         }{%
        \isChicagoStyle{%
        \AuthorCitation{Lastname, Firstname, Firstname Lastname, and Firstname Lastname.}
        }{
        \AuthorCitation{Vannah, S.; Stiehl, I.D.; Gleiser, M.}
        }
}

\address{%
$^{1}$ \quad Atmospheric and Environmental Research, Inc., Lexington, MA 02421, USA\\
$^{2}$ \quad Department of Physics and Astronomy, Dartmouth College, Hanover, NH 03755, USA; {mgleiser@dartmouth.edu (M.G.)}
\\
$^{3}$ \quad Department of Physical Sciences, Earth and Environment, University of Siena, 53100 Siena, {} 
 Italy}

\corres{Correspondence: svannah@aer.com}

\abstract{In the past, measures of the ``Earth-likeness'' of exoplanets have been qualitative, considering an abiotic Earth, or requiring discretionary choices of what parameters make a planet Earth-like. With the advent of high-resolution exoplanet spectroscopy, there is a growing need for a method of quantifying the Earth-likeness of a planet that addresses these issues while making use of the data available from modern telescope missions. In this work, we introduce an informational--entropic metric that makes use of the spectrum of an exoplanet to directly quantify how Earth-like the planet is. To illustrate our method, we generate simulated transmission spectra of a series of Earth-like and super-Earth exoplanets, as well as an exoJupiter and several gas giant exoplanets. As a proof of concept, we demonstrate the ability of the information metric to evaluate how similar a planet is to Earth, making it a powerful tool in the search for a candidate Earth 2.0.}


\begin{document}

\section{Introduction}

The search for Earth-like, potentially inhabited planets has long been a key driver of discovery in astronomy. Current and near-future spectral observations are ushering this search into a new era. In~the past, limited resolution and spectral coverage meant that exoplanets---especially rocky, potentially Earth-like planets---were assessed by a combination of inferred quantities (such as mass, radius, and~distance from host star) and computer modeling. However, future missions such as the Atmospheric Remote-sensing Infrared Exoplanet Large-survey (ARIEL) \citep{tinetti2018chemical}, Large Interferometer For Exoplanets \mbox{(LIFE) \cite{quanz2022}},  the~Habitable Worlds Observatory (HWO) \cite{decadal, Harada2024}, Earth-2.0~\cite{ye2022china}, and~the European Extremely Large Telescope (E-ELT) \citep{ramsay2020eso} are focused partially or wholly on the search for an Earth-analog planet and include high-resolution, wide-passband spectroscopy, as~well as advancements in telescope technology such as adaptive optics and novel observational techniques. As~such, they promise to revolutionize the characterization of Earth-like exoplanets, by~providing, for~the first time, observational information about the chemical composition and structure of planets' atmospheres. In~fact, spectroscopy has already begun to revolutionize exoplanet characterization: JWST observations have been used to identify carbon dioxide, water, sulfur dioxide, and~sulfur monoxide in the atmosphere of gas giant WASP-39~b~\citep{rustamkulov2023early, ahrer2023early, alderson2023early, feinstein2023early}, water in the gas giant WASP-96 b \citep{Pontoppidan_2022}, and~carbon dioxide and methane in K2-18 b \citep{Madhusudhan_2020}. Assessments of the Earth-likeness of exoplanets should take this new, invaluable source of information into account. Furthermore, we will argue that an assessment of the Earth-likeness of an exoplanet should be holistic, including not just the intrinsic astronomical properties of the planet in isolation but~the atmospheric impacts of the life it hosts, if~any. As~we will show, information theory can provide a quantitative metric to compare the different observed exoplanets and effectively assess their Earth-likeness. 

The term ``Earth-like'' is used in two distinct ways in the exoplanet community: as a qualitative description of a planet, or~as a quantitative metric. Qualitative definitions vary across the astronomical community. The~term is used most commonly to group rocky worlds with radii $0.5R_\oplus \lesssim R_p \lesssim 1.5R_\oplus$, where $R_\oplus$ is Earth's radius, and~$R_p$ is the exoplanet's. An~extra condition that is sometimes applied is that an Earth-like world must be in the habitable zone of its host star. Although~useful as a first step, these methods offer an incomplete description of what it truly means for a planet to be similar to a living world like Earth. The current quantitative approaches to quantify the Earth-likeness or habitability of an exoplanet can be categorized into two classes: metrics relying on a short list of parameters describing the planet and approaches that use machine learning to manage a longer list of parameters. We briefly review these approaches~here. 

The Earth Similarity Index (ESI) \cite{schulzemakuch2011} and related expansions~\cite{Kashyap2017, jagadeesh2018earthsimilarityindexhabitability} are perhaps the most widely used of the short-list planetary comparison metrics, for~example, in NASA's Planetary Habitability Laboratory. These metrics compute a weighted sum of the difference between the radius of the planet, surface gravity, distance from host star, and~surface temperature of the planet compared to those of Earth. Similar measures such as the Constant Elasticity Earth Similarity Approach (CEESA) \cite{basak2020} and Cobb--Douglas habitability production function (CD-HPF) \cite{bora2016} use alternative calculations with different parameters (planetary radius, density, surface temperature, escape velocity, and~eccentricity for CEESA, and radius, density, escape velocity and surface temperature for CD-HPF). Additionally, several metrics have been proposed that use parameters describing the habitability of a planet, rather than the intrinsic Earth-likeness. The~Planetary Habitability Index (PHI), for~instance, is a complement to the ESI meant to assess the likelihood that a planet is actually capable of hosting life and depends on the presence of a stable substrate, the energy available (largely from sunlight and chemistry), a~chemical makeup that can form the polymeric chemistry necessary for life, and~the presence of liquids that can be used as a solvent~\cite{schulzemakuch2011}. Another measure, the~Biological Complexity Index (BCI) considers the presence of a stable substrate, surface temperature, age of the planet, geological activity, and~energy availability~\cite{challe5010159}.

A second class of quantitative Earth-likeness metrics uses machine learning to incorporate a larger list of planetary parameters. \citet{Sarkar2021} use an unsupervised learning approach to search for anomalies in exoplanet features. Depending on the dataset used, these features may include parameters related to the mass, temperature, and~size of both the planet and the star, the~distance between the two, or~orbital parameters. \citet{saha2018} use a supervised approach to assess~habitability.

While both the qualitative and quantitative methods allow for the categorization of Earth-like planets, most share a common limitation: they require a discretionary, {}{a priori} 
 choice of what parameters make a planet ``Earth-like''. ESI and related measures, for~example, choose a short list of parameters to describe the Earth-likeness of a planet. This parametrization becomes especially problematic when searching for potentially habitable Earth-like planets due to the unsettled debate of what planets qualify as ``habitable'' \cite{seager2014future, seager2013exoplanet, Schwieterman2018}. We propose that an improved method to assess the Earth-likeness of exoplanets should be agnostic of which parameters describe the Earth-likeness of the~planet.


Furthermore, the~previous methods have assessed either the habitability \textit{or} the Earth-likeness of a planet. We argue that these two traits are not separable. In~fact, Earth itself was and continues to be measurably changed by the presence of life. The~detection of these so-called biosignatures in an exoplanetary spectrum would be a strong indicator that the planet hosts life \citep{seager2005}. Since our proposed method uses the spectra of planets, the~metric is sensitive to the presence (or absence) of life on the planet. Potential biosignatures include the vegetative red edge~\cite{omalley2018, sagan1993, seager2005}, metabolic by-products in out-of-equilibrium chemistry such as combinations of methane and ozone, or~methane and O$_2$. Many of these molecular features will be detectable with the upcoming telescope missions under certain circumstances (e.g., low clouds) \cite{Thompson2022, Tinetti2021, udry2014exoplanet, wang2018}.  This makes our method well timed as~we enter an era of higher-precision spectroscopic exoplanet~characterization.


In this work, we propose an information theory-based method designed to be applied to transit spectroscopy. Like the ESI and its expansions, our method is a quantitative metric. Like atmospheric retrievals, our method can identify the Earth-likeness of an exoplanet's atmospheric composition. And~like molecular species identification, our method is sensitive to the presence of biosignature gases. The~method introduced here is a companion to that of \citet{vannah2024}. While the method introduced in that work uses the information content of exoplanet atmospheres as a function of wavelength to analyze specific biosignature gases and habitability indicators, here we develop a method which uses the information content of a wide bandpass of the spectrum to assess the Earth-likeness---including the possible presence of life---of a particular planetary~spectrum. 

This is the main goal of the informational measure we propose here. It can be used to search for different kinds of exoplanets---not just Earth-like ones---focusing on their spectral signature. Within~our approach, an~Earth-like planet would be one that has a spectral signature with informational content (as defined in Section~\ref{sec:Info}) close to Earth's. An~Earth clone would be one with a spectral signature identical to Earth's--and thus with identical informational content. Since a true Earth clone is not realistic---due to instrumental limits and because no two planets may be perfectly alike---the difference in the information content between two planets may never be zero. Rather, two planets are most similar when their difference in information content is minimized. Our method provides a holistic, quantitative measure of the Earth-likeness of an exoplanet that complements the data already known from combining transit and Doppler methods that can furnish the radius, mass, and~distance from the host~star. 

By quantifying the information content in the exoplanetary spectrum, we are able to identify---for spectral resolution with sufficiently low noise---spectra similar to those of Earth: the smaller the difference in information content, the~more similar the spectra and thus the more Earth-like the exoplanet. Since we are still in the early days of acquiring high-quality spectral data (we will specify what high quality means for our method), our goal here is to offer a proof of concept, using our method to compare simulated exoplanets to a simulated Earth spectrum. We show that our information measure efficiently differentiates between Earth-like and Jupiter-like planets. We also demonstrate that our method is sensitive to variations in physical parameters that affect a planet's spectral~signature. 


Section~\ref{sec:Info} (Information Measure) describes the information theory metric we employ to differentiate planets. Section~\ref{sec:data} (Data) describes the simulations used to obtain data for this analysis. In~Section~\ref{sec:Res1} (Results 1), we show how our method is affected by varying the physical parameters for a sample Jupiter-like planet. This demonstrates the ability of our method to differentiate between planet types and to identify changes in planetary features. In~Section~\ref{sec:Res2} (Results 2), we show the results for a series of simulations of observed  gas giant and  rocky exoplanets. We compare these exoplanets with the simulated spectra of Earth, Jupiter, and~a hot Jupiter clone to show how our method is able to discriminate between types of planets with realistic, diverse variations in planetary features. In {\mbox{Section \ref{sec:Conclusions}} (Discussion and Conclusions)} 
we summarize our results and expand on how the method may be used to identify the biosignatures associated with inhabited planets. Finally, we present an error analysis in the 
{}{Appendix \ref{sec:appendix}}
.

\section{Information~Measure}\label{sec:Info}

{}{The} 
 application of an informational entropic method to transmission spectra was introduced in a companion paper, \citet{vannah2024}, and is summarized here for reference. 
In this work, we quantify the dissimilarity between two planetary spectra as the difference between the information contents contained in the two spectra, or~more precisely, the~distance in the information space. We treat the spectra as probability {}{distributions,} 
 \begin{equation}
    p_\nu = \frac{D_\nu}{\sum_\nu D_\nu},
\end{equation}
for $D_\nu$, the transit depth of the spectrum at a particular wavenumber, $\nu$. This takes the form of a modal fraction, probability distributions introduced \mbox{by \citet{Gleiser2012}} to quantify the configurational entropy of spatially localized configurations of scalar fields. The~information content of a modal fraction can be quantified through its Shannon \mbox{entropy \citep{Shannon1948}} $H$, given by
\begin{equation}
    H = -\sum_\nu p_\nu log(p_\nu).
\end{equation}

Similarly, the~Shannon entropy of a planetary spectrum quantifies the syntactic (non-meaningful, as~opposed to semantic) information content of the spectrum. The~information required to discriminate between two spectra is given by the Jensen--Shannon \mbox{Divergence \citep{lin1991divergence}}, 
\begin{equation}
    \djs(p||q) =\frac{1}{2}\dkl(p||r) +\frac{1}{2}\dkl(q||r),
    \label{eq:djs}
\end{equation}
for $r=\frac{1}{2}(p+q)$ and $\dkl$, the Kullback--Leibler Divergence \citep{Kullback:1951}, 
\begin{equation}\label{eq:djs}
    \dkl(p||q) = \sum_\nu p_\nu log (\frac{p_\nu}{q_\nu}). 
\end{equation}

For an Earth spectrum with modal fraction $p_\nu$, $\djs (p||q)$ is a metric representing the amount of information lost by replacing the Earth spectrum with an exoplanet spectrum with modal fraction, $q_\nu$. A~smaller information loss indicates more similar spectra; a $\djs$ of zero indicates identical distributions. Higher $\djs$ represents more information loss (and therefore less similarity) between two spectra. 

In log base 2, the~information measured in $\djs$ is in bits. A~bit is the amount of information contained in an event with probability $\frac{1}{2}$, such as flipping a coin. Each ``event'' (data point) in the spectral distributions $p$ and $q$ has a probability far smaller than $\frac{1}{2}$, so the information encoded in each data point is far less than one bit. In~this work, we use the natural logarithm so that  $ \djs$ is given in units of nats.  We emphasize that $\djs$ is a purely comparative, rather than absolute, measure, as~the amount of information contained in a spectrum depends on the resolution and noise of the specific datasets. However, for~a collection of spectra with similar noise and resolution, $\djs$ can be used to quantitatively compare and group planets with similar spectral features.

While there are simpler approaches to compare planetary spectra---for example, simply taking the difference $[p_\nu - q_\nu]$---our method is, critically, a~distance metric. Distance metrics are commonly used in statistics to create a metric space where two points in the space (here, the~two points correspond to two planets) have some quantifiable distance between them. Like a distance measure in regular geometry, metrics measure distances that are always positive (or zero, for~the distance between a point and itself), symmetric (the distance in information space between, say, Earth and Mars, is the same as the distance between Mars and Earth), and~obey the triangle~inequality. 

\newpage
In contrast to previous methods of exoplanet characterization, $\djs$ requires no prior knowledge to interpret the spectrum. This is critical: biosignatures or signs of habitability may resemble Earth's atmosphere in ways we cannot predict {}{a priori}. There may also be examples of ``life as we don't know it'', producing biomarkers dissimilar from those we may know to look for. Relying on our incomplete understanding of the diversity of scenarios that may host life could cause us to miss inhabited planets. $\djs$ provides a holistic, quantitative measure of the Earth-likeness of an exoplanet. Furthermore, $\djs$ scales in a predictable manner with noise. This calculation in shown in Appendix \ref{sec:appendix}.

We note that $\djs$ alone is not able to isolate biosignatures, as it measures the difference in information content of the full spectrum rather than individual absorption lines. (We can think of it as a global quantity obtained from a given spectrum, like the area between two points under a curve obtained from its integral.) The identification of biosignatures requires some form of input knowledge (e.g., specific compounds related to biotic activity), while $\djs$ requires none. In~a companion paper \citep{vannah2024}, we showed how the $\djs$-density per wave number can be used to isolate specific biosignatures in planetary spectra with limited input knowledge (compare to, for~instance, Equation (\ref{eq:djs}) or Equation~(14) in~\cite{Gleiser2012a}). Together, these two uses of $\djs$ allow for both the identification of a Earth-like planets (this paper) and for a detailed characterization of an exoplanet's atmospheric composition. The~global $\djs$ studied here may be used to search for exoplanets with the potential to host life, while the $\djs$-density may identify which of those exoplanets may actually be~inhabited.

\section{Data} \label{sec:data}
As a test of our measure, we demonstrate that $\djs$ can identify different classes of exoplanet simulations using only their spectra. To~illustrate this point, we use the radiative transfer code \classoption{{}{Exo-Transmit}
} ({}{\url{https://github.com/elizakempton/Exo_Transmit}} (accessed on 24 January 2022)
) for our simulations; further details on the mechanics of the simulation may be found in \mbox{\citet{kempton_exo-transmit_2017}}. The~simulations are governed by seven parameters: equilibrium temperature used for a temperature-pressure profile, atmospheric equation of state, planetary surface gravity, planetary radius, stellar radius, a~parameter controlling the cloud top pressure, and~a parameter controlling the strength of Rayleigh scattering. For~simplicity, we use an isothermal temperature-pressure profile, as~transmission spectra are principally absorption spectra and thus minimally sensitive to temperature gradients. The~simulation includes CH$_4$, CO$_2$, CO, H$_2$O, NH$_3$, O$_2$, O$_3$, C$_2$H$_2$, C$_2$H$_4$, C$_2$H$_6$, H$_2$CO, H$_2$S, HCl, HCN, HF, MgH, N$_2$, NO, NO$_2$, OCS, OH, PH$_3$, SH, SiH, SO$_2$, TiO, VO, Na, and~K. The~simulation uses a spectral resolution $\frac{\Delta \lambda}{\lambda} = 10^3$, and~has a wavelength range from \mbox{0.3 $\upmu$m} to 30 $\upmu$m. The~simulated spectra used in this work are shown in Appendix \ref{sec:sims}. 

For all of our simulations, we choose to set a fixed value for the Rayleigh scattering parameter while allowing the code to calculate the cloud top pressure. We use the literature values of each of the parameters to simulate an Earth spectrum and a Jupiter spectrum, using equilibrium chemistry for simplicity. We create a series of Jupiter clones with each individual parameter varied in isolation to demonstrate their effects on $\djs$, shown \mbox{in Section~\ref{sec:Res1}}.

We also simulate realistic  gas giants and rocky exoplanets to demonstrate the ability of $\djs$ to differentiate between planet types. For~each exoplanet class, we create ten planet simulations: six from observed exoplanets using parameter values from the literature, and~four artificial planets designed to explore the parameter space. The~parameters used to create these exoplanets are shown in Table~\ref{tab:params}. For~comparison with the exoplanets, we simulate a Jupiter spectrum, an~Earth spectrum, and~the spectrum of a Jupiter clone with the temperature raised to 1200 K. These results are shown in
Section~\ref{sec:Res2}. While Exo-Transmit cannot be directly validated against Earth or Jupiter transmission spectra, similar forward models have been validated through satellite observations of Earth by comparison with infrared, near-infrared, and~visible transit spectra (using solar occultations) and Earthshine measurements \citep{Lustig-Yaeger_2023, macdonald2019, Kaltenegger_2009, robinson2011, Kaltenegger_2007}.


\begin{table}[H]
\footnotesize
\caption{{}{The} 
 parameters used to generate realistic exoplanet simulations. The gas giant simulations are shown in the top half of the table, while the  rocky planets are shown in the bottom half. The~two planetary classes are separated by a horizontal line. 
{}{$^a$} 
 Sources: \citet{montanes-rodriguez_jupiter_2015, sato_jupiters_1979}{}{;} 
 {}{$^b$} Source: \citet{liu_high_2014}{}{;} {}{$^c$} Source: \citet{hartman_hat-p-12b_2009}{}{;} {}{$^d$} Source: \citet{boyajian_stellar_2015}{}{;} \mbox{{}{$^e$} Source: \citet{del_burgo_accurate_2016}}{}{;} {}{$^f$} Source: \citet{gillon_discovery_2009}{}{;} {}{$^g$} Source: \citet{faedi_wasp-39b_2011}{}{;} \mbox{{}{$^h$} Source: \citet{kempton_exo-transmit_2017}}{}{;} {}{$^i$} Sources: \citet{lin_high-resolution_2020, barnes_habitability_2016}{}{;} {}{$^j$} Sources: \citet{lin_high-resolution_2020, delrez_early_2018}{}{;} {}{$^k$} Source: \citet{pinamonti_hades_2018}{}{;} {}{$^l$} Source: \mbox{\citet{anglada-escude_dynamically-packed_2013}}{}{;} {}{$^m$} Source: \citet{bauer_carmenes_2020}{}{;} {}{$^n$} Source: \citet{hidalgo_three_2020}.
\label{tab:params}}

\begin{adjustwidth}{-\extralength}{0cm}
\centering 
\begin{tabularx}{\fulllength}{lLlLLLL}
\toprule
\textbf{Name} & \textbf{Equilib. Temp. (K)} & \textbf{Equation  of  State} & \textbf{Surface  Gravity (g)} & \textbf{Planet  Radius (m)} & \textbf{Stellar  Radius (m)} & \textbf{Rayleigh   Scattering  Factor} \\
\midrule
Jupiter {}{$^a$} & 300 & 1{}{X} 
 & 24.79 & 6.99 $\times$ 10$^7$ &  6.96 $\times$ 10$^8$  & 10 \\
HAT-P-1b {}{$^b$} & 1300 & 5{}{X}, graphite rainout & 7.5  & 9.44 $\times$ 10$^7$ & 8.17 $\times$ 10$^8$ & 10 \\
HAT-P-12b {}{$^c$} & 1000 & 1{}{X} & 5.6 & 6.86 $\times$ 10$^7$ & 4.87 $\times$ 10$^8$  & 200 \\
HD 189733b {}{$^d$} & 1200 & 1{}{X} & 21.4 & 8.15 $\times$ 10$^7$ & 5.60 $\times$ 10$^8$  & 500 \\
HD 209458b {}{$^e$} & 1500 & 0.1{}{X} & 9.4 & 9.72 $\times$ 10$^7$ & 8.35 $\times$ 10$^8$  & 10 \\
WASP-6b {}{$^f$} & 1200 & 1{}{X} & 8.7 & 8.72 $\times$ 10$^7$ & 6.05 $\times$ 10$^8$ & 1000 \\
WASP-39b {}{$^g$} & 1100 & 1{}{X} & 4.1 & 9.08 $\times$ 10$^7$ & 6.23 $\times$ 10$^8$ & 1 \\
Gas giant 1 & 1400 & 5{}{X}, graphite rainout & 12.8 &  8.45 $\times$ 10$^7$ & 5.02 $\times$ 10$^8$ & 1000 \\
Gas giant 2 & 700 & 1{}{X} & 20.1 & 6.27 $\times$ 10$^7$ & 8.95 $\times$ 10$^8$ & 100 \\
Gas giant 3 & 1000 & 1{}{X}, 0.2 C/O ratio & 8.8 & 9.23 $\times$ 10$^7$ & 6.48 $\times$ 10$^8$ & 10 \\
Gas giant 4 & 1300 & 1{}{X}, 0.8 C/O ratio & 17.2 & 6.90 $\times$ 10$^7$ & 9.23 $\times$ 10$^8$ & 1 \\
\midrule
Earth {}{$^h$} & 300 & 1{}{X} & 9.8 & 6.37 $\times$ 10$^6$ & 6.96 $\times$ 10$^8$ & 1 \\
Proxima b {}{$^i$} & 300 & 1{}{X} & 10.9 & 6.82 $\times$ 10$^6$ & 9.82 $\times$ 10$^7$ & 1 \\
TRAPPIST-1e {}{$^j$} & 300 & 1{}{X} & 7.2 & 5.85 $\times$ 10$^6$ & 8.15 $\times$ 10$^7$ & 1 \\
GJ 15 Ab {}{$^k$} & 300 & 0.1{}{X} & 12.4 & 9.88 $\times$ 10$^6$ & 2.85 $\times$ 10$^8$ & 10 \\
GJ 667 Cc {}{$^l$} & 300 & 0.1{}{X} & 15.7 & 9.81 $\times$ 10$^6$ & 2.92 $\times$ 10$^8$ & 1 \\
CD Cet b {}{$^m$} & 500 & 1{}{X} & 11.7 & 1.16 $\times$ 10$^7$ & 1.18 $\times$ 10$^8$ & 0 \\
EPIC 24983012b {}{$^n$} & 1500 & 1{}{X} & 22.6 & 1.24 $\times$ 10$^7$ & 1.19 $\times$ 10$^9$ & 1 \\
Rocky planet 1& 700 & 1{}{X} & 14.8 & 9.87 $\times$ 10$^6$ & 8.80 $\times$ 10$^8$ & 10 \\
Rocky planet 2 & 400 & 5{}{X} & 10.4 & 6.02 $\times$ 10$^6$ & 4.53 $\times$ 10$^8$ & 1 \\
Rocky planet 3 & 300 & 1{}{X} & 8.4 & 7.12 $\times$ 10$^6$ & 9.25 $\times$ 10$^8$ & 1000 \\
Rocky planet 4 & 1000 & 10{}{X} & 12.8 & 8.54 $\times$ 10$^6$ & 1.02 $\times$ 10$^9$ & 1 \\
\bottomrule
\end{tabularx}
\end{adjustwidth}
\end{table}
\unskip

\section{Results 1: Comparing Spectra by Changing Physical~Parameters}
\label{sec:Res1}
\begin{figure}[H]

 \includegraphics[width=\textwidth]{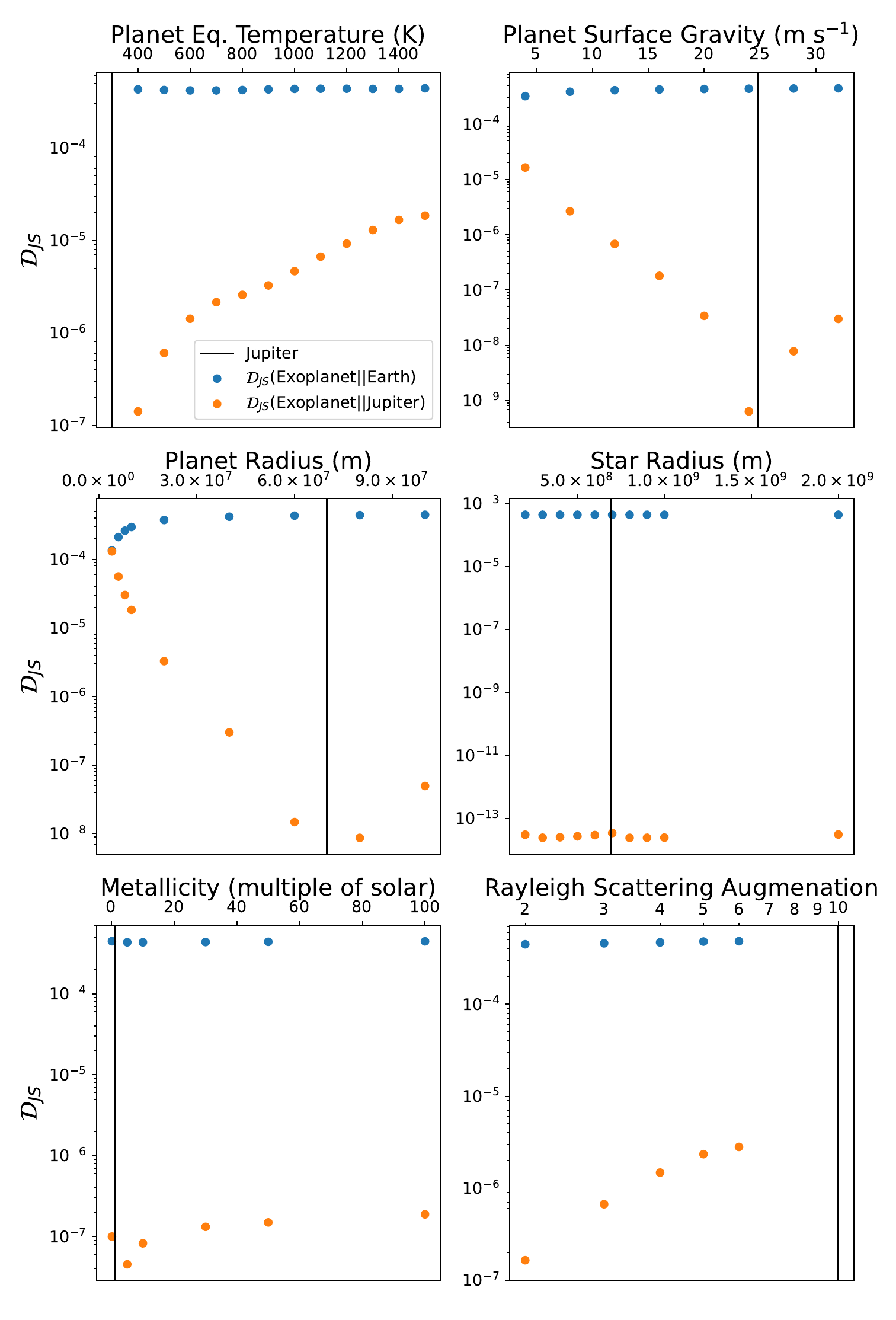}

	\caption{$\djs$ {}{of} 
 a simulated, Jupiter-like exoplanet compared to Earth (blue dots) and to Jupiter (orange dots). Each plot represents the variation of one parameter. The~vertical black line indicates the Jupiter value of the~parameter.}
	\label{fig:params}
\end{figure}

We begin by testing our information measure in a simple scenario, varying the physical parameters of the simulated planets to assess how well our our informational measure picks up on these changes when compared to Earth and Jupiter. Figure~\ref{fig:params} shows how varying each of the input parameters in isolation affects the $\djs$ of a model Jupiter-like planet relative to Earth and to Jupiter. Variations in the values of each of the planetary parameters away from their Jupiter values (vertical black line) incrementally increase the $\djs$ relative to Jupiter (orange points), displaying the growing dissimilarity between the two planets. This demonstrates that $\djs$ is sensitive to variations in planetary features. The~$\djs$ relative to Earth (blue points) remains higher than the $\djs$ relative to Jupiter (orange points), confirming that small changes in the planetary parameters do not make the sample planet appear~Earth-like.

Each of the plots tells a story illustrating the ability of $\djs$ to pick up on the underlying physics. Using the modal fraction ensures that the information content is not dependent on the continuum flux,  only on the relative strength and shape of the absorption lines. For~the equilibrium temperature, surface gravity, planet radius, and stellar radius---the top \mbox{two rows} of plots---varying the parameters varies the strength of the absorption lines in the modal fraction. For~example, the~scale height, $H$, of~the planetary atmosphere is linearly dependent on the temperature, $T$, given by
\begin{equation}
    H = \frac{k_B}{\mu_m g} T,
\end{equation}
for $k_B$ the Boltzmann constant, $\mu_m$ the mean molecular mass, and~$g$ the surface gravity on the planet. Increasing the scale height increases the size of the light-absorbing atmosphere. Therefore, the~absorption line strength grows with temperature. As~the strength of the exoplanet absorption lines diverges from the strength of the absorption lines in the Jupiter spectrum, the~information contained in the two spectra also diverge. Vice~versa, decreasing the temperature  decreases the strength of the absorption lines. This results in a valley in the temperature plot (top left), where the lowest point in $\djs$---highest similarity between the two planets---sits nearest the Jupiter temperature (black line), while increasing incrementally as the temperature increases. Since, except~for metallicity and Rayleigh scattering, changing the parameters away from the Jupiter value impacts the strength of the absorption lines, we see a valley occurring in nearly all plots. This illustrates the reliability of our approach to distinguish between different planetary properties. We note that the stellar radius plot in the middle right indicates that small variations in the stellar radius make little impact compared to variations in the other parameters. This is represented by the extremely low (of order $10^{-13}$) values of $\djs$ relative to Jupiter for the range of parameter values we~analyzed. 

Increasing the strength of Rayleigh scattering causes information divergence not by impacting the strength of the absorption lines but~by creating a spectral tilt in the near-infrared \citep{des2008rayleigh}. This reduces $\djs$ at the low wavelength absorption lines in the spectrum. Therefore, just as with the other parameters, the~values of the Rayleigh scattering factor closest to Jupiter's have the lowest $\djs$. Similarly, changing the metallicity impacts the relative strength of absorption lines for different molecules. As~a consequence, metallicities closest to Jupiter have the lowest $\djs$, although~the differences are relatively small (smaller than an order of magnitude) compared to other physical~parameters.

The second key result from the parameter variation analysis is that the values of $\djs$ comparing the modified Jupiters to Earth (blue dots) are consistently higher than those comparing to Jupiter. This indicates that our method is able to distinguish Earth-like and Jupiter-like planets for a wide range of parameters. The~only exception occurs at the smallest planet radius in the middle left plot. As~the planetary radius is decreased to even smaller values than the Earth radius, the~$\djs$ compared to Jupiter and to Earth approach each other. This does not indicate that the planet is similar to Earth. Rather, it indicates that the distance in information space between the planet and Jupiter is similar to the distance between the planet and Earth, albeit in different directions. For~the planet and Earth to be similar, the~$\djs$ between the two would need to be small. How small would be hard to determine, unless~we had a large sample of exoplanets that included ones that have spectra that are truly similar to Earth's. We will be more specific about this when we discuss our results~below.

\section{Results 2: Differentiating Between Exoplanet Types with the Information~Metric}\label{sec:Res2}
In the previous section, we investigated the effects of changing a single physical parameter at a time as a first illustration of using $\djs$ as a discriminator of specific exoplanet properties when compared to a chosen baseline planet. Of~course, in~reality, the physical parameters that determine an exoplanet spectrum are often interdependent. Moving thus toward more concrete situations, in~this section, we simulate the spectra of several observed rocky planets and gas giants to show that $\djs$ can indeed differentiate between planetary classes. We compare the two classes of planets to simulated Earth and Jupiter spectra, as~well as to the spectrum of a Jupiter clone with the equilibrium temperature increased to 1200 K. Our results show that $\djs$ can identify which of the simulated worlds is closest to Jupiter or to~Earth. 

\begin{figure}[H]

    \includegraphics[scale=0.8]{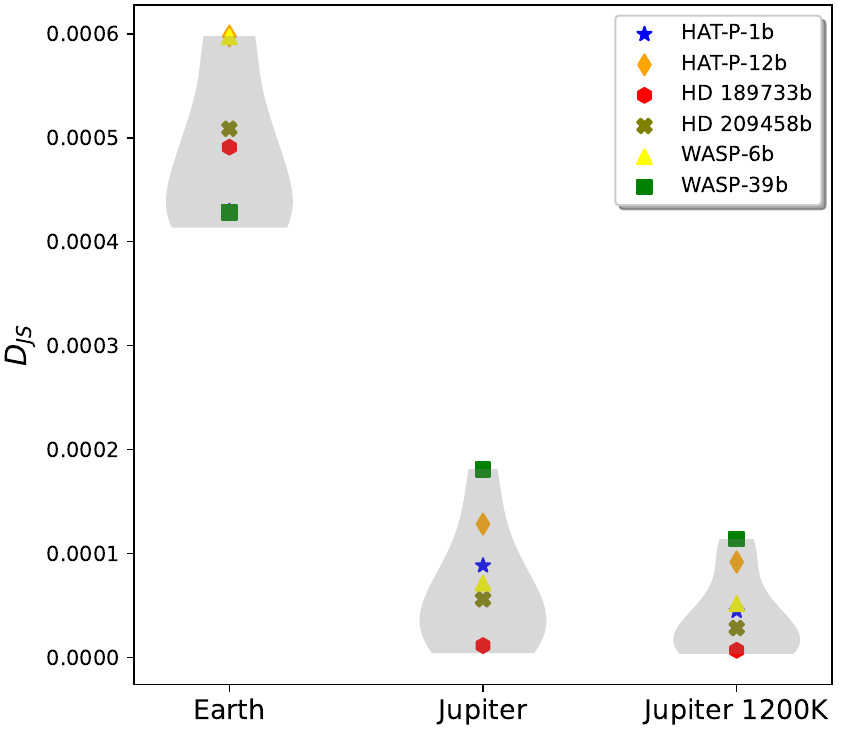}
    \caption{{Violin}  plot showing the $\djs$ distributions comparing the simulated gas giants to Earth (\textbf{left}), to~Jupiter (\textbf{center}), and~to a Jupiter clone with the temperature raised to 1200 K (\textbf{right}). The~width of the violin plot reads like a histogram, indicating the distribution of planets clustered around the mean $\djs$ value. The widest areas of the violin plot indicate the most common $\djs$ value for a particular distribution. The $\djs$ from six observed exoplanets are labeled with colored shapes; the remaining points in the $\djs$ distribution are from four exoplanet simulations. The physical parameters for the exoplanets are listed in Table~\ref{tab:params}.}
    \label{fig:hotJups}
\end{figure}

In Figure~\ref{fig:hotJups}, we find that the $\djs$ distribution comparing the six gas giants to the two Jupiter-like planets do not overlap with the $\djs$ distribution comparing them to Earth. This illustrates that $\djs$ is able to distinguish between Jupiter-like and Earth-like planets. There is, however, an~overlap between the two distributions comparing the gas giants to Jupiter and to a 1200 K Jupiter. This is due to the similarity of the comparison planets. Still, the~mean (bulge in the gray violin plots) of the 1200 K Jupiter $\djs$ distribution is the lower of the two distributions, indicating that $\djs$ is able to identify the temperature similarity of the hot Jupiters and the hotter, 1200 K Jupiter clone. We also show the results of specific comparisons for the six gas giants, each labeled by a different colored shape. For~example, planet WASP-39b (identified by a green square) is clearly a hot Jupiter, most similar to the simulated Jupiter at~1200 K.

\begin{figure}[H]

\includegraphics[scale=0.78]{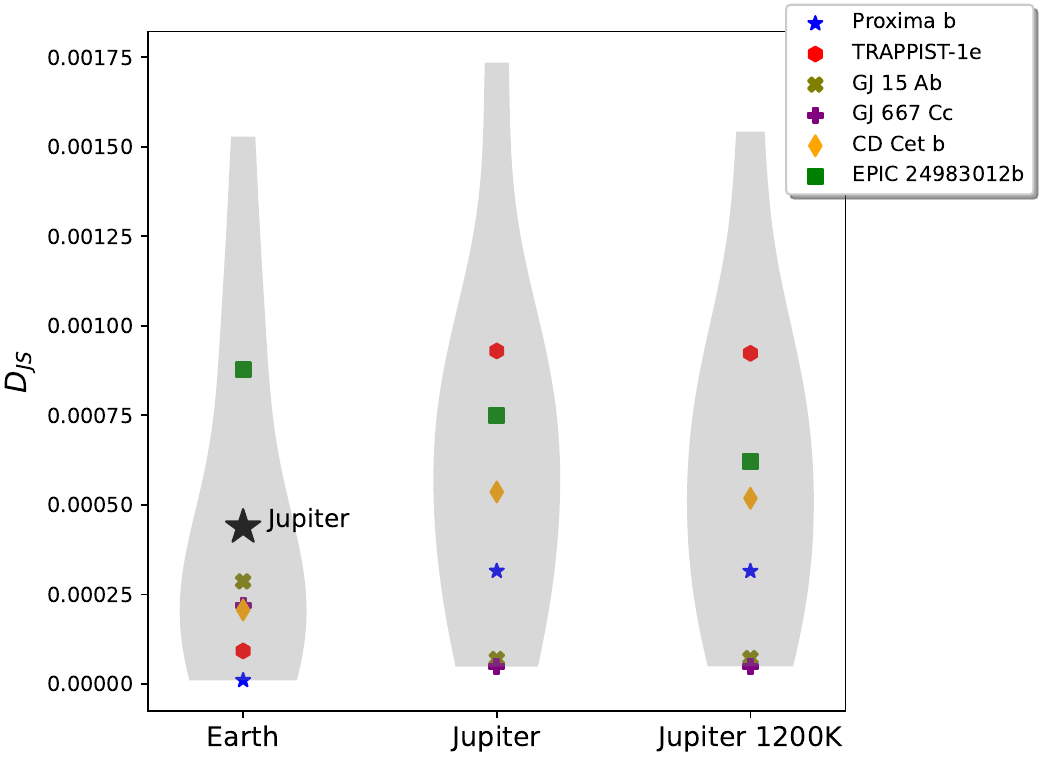}

	\caption{{Violin} plot showing the $\djs$ distributions comparing the simulated  rocky planets to Earth (\textbf{left}), to~Jupiter (\textbf{center}), and~to a Jupiter clone with the temperature raised to 1200 K (\textbf{right}). The~width of the violin plot reads like a histogram, indicating the distribution of planets clustered around the mean $\djs$ value. The~$\djs$ from observed exoplanets are labeled with colored shapes; the remaining points in the $\djs$ distribution are from four exoplanet simulations. The physical parameters for the exoplanets are listed in Table~\ref{tab:params}. We include the $\djs$ comparing Jupiter and Earth in the leftmost violin plot for reference. Note that the position of Jupiter on this plot indicates that all the  rocky planets but EPIC 24983012 are more Earth-like than~Jupiter.}
	\label{fig:supEarths}
\end{figure}

The rocky planet $\djs$ distributions in Figure~\ref{fig:supEarths} show more general overlap, although~different exoplanets (colored shapes) are still clearly distinguished when compared to Earth (left) and to the two Jupiters (center and right). The~mean $\djs$ (the bulge in each plot) comparing the  rocky planets to Earth is lower than comparing them to Jupiter or to a \mbox{1200 K} Jupiter, although~the distributions are less distinct than those comparing gas giants in Figure~\ref{fig:hotJups}. Still, when compared to Earth (left plot), with~the exception of exoplanet EPIC 24983012b, all others have $\djs$ substantially lower than the gas giants of Figure~\ref{fig:hotJups}, indicating that the method distinguishes between the two classes. In~particular, we note how Proxima b (blue star) is the closest exoplanet to Earth in this sample. In~contrast, EPIC 24983012b has an equilibrium temperature 1200 K hotter than Earth and a surface gravity more than twice that of Earth. The~high $\djs$ between this exoplanet and Earth indicates that $\djs$ is able to pick up on these physical differences. EPIC 24983012b may be a super-Earth, but~it is definitely not Earth-like. 
In general, super-Earths have larger radii and typically higher temperatures than Earth, moving them closer to the gas giants. Our results are consistent with~this.

\vspace{-6pt}

Within the accuracy of our results, we can now propose a preliminary criterion based on $\djs$ to distinguish between exoplanets. In~Figures~\ref{fig:hotJups} and \ref{fig:supEarths}, the~leftmost violin plots depict comparisons with Earth: in Figure~\ref{fig:hotJups} between Earth and gas giants, and~in Figure~\ref{fig:supEarths} between Earth and super-Earths or Earth-like worlds. We note that the results clearly depend on the sample of planets we are using to compare, and~on the quality of the signal-to-noise ratio (see Appendix \ref{sec:appendix}). With~those caveats, from~Figure~\ref{fig:hotJups} and the comparison with Earth, if~$\djs({\rm exoplanet||Earth}) > 3\times 10^{-4}$, the exoplanet is most probably a gas giant. From~Figure~\ref{fig:supEarths}, the~exoplanets that have spectra closest to Earth's would have $\djs({\rm exoplanet||Earth}) \leq 2\times 10^{-4}$, which from our sample would include the two most Earth-like exoplanets, TRAPPIST-1e and Proxima~b.

\section{Discussion {}{and} 
 Conclusions}\label{sec:Conclusions}
In this work, we have introduced a new metric that makes use of modern spectroscopic data to assess the Earth-likeness of planetary spectra. The method can also be applied to compare a planetary spectra to another planet---say, Jupiter, as~in our work---providing an agnostic measure based on any reference spectrum, given sufficient accuracy. Our metric has several key strengths: It is a quantitative assessment that can be used to rank the Earth-likeness of different exoplanets, placing it into a well-established field of indices assessing similarity to Earth. Unlike these indices, it does not require an  {}{a priori} choice of what parameters make a planet Earth-like, making our method independent of the unsettled debate of what makes a planet Earth-like or habitable. Furthermore, the~accuracy of our information  metric for assessing the Earth-likeness of exoplanets will increase with data availability from future and upcoming high-resolution wide-passband spectroscopic missions. Finally, our quantitative method is holistic, as~it does not separate Earth from its~life. 


This work is a companion to \citet{vannah2024}, which introduced a wavelength-specific comparison (a $\djs$-wavelength density) between two spectra to search for biosignatures. The~difference in information content between an Earth-as-an-exoplanet spectrum and the spectrum of an exoplanet at a wavelength corresponding to a potential biosignature gas can indicate the likelihood that the planet hosts Earth-like life. Similarly, as~introduced in this work, the~total information content relative to Earth contained in a spectral window can indicate how Earth-like an exoplanet is. While the method introduced \mbox{in \citet{vannah2024}} can be used to search for specific gasses potentially associated with biosignatures, the~metric introduced here can be used to search for potentially habitable planets in a global ``Earth-like''~sense.

\citet{Stephens2020} and other works cited therein show that $\djs$ is sensitive to patterns in the data. The~better the signal-to-noise ratio and the wavelength resolution of the spectrum, the~more efficient the method. This dependence is explored further in Appendix~\ref{sec:appendix}. Similar approaches using the information entropic content of a spectrum or a field have been shown to be effective in a wide variety of astrophysical, cosmological, and~high-energy physics scenarios: \citep{thakur2020configurational,  bernardini2019, Gleiser2012, Gleiser2012a, Gleiser2013, Gleiser2018, bernardini2017, Stephens2020, braga2017} is an incomplete list. In~this work, we use $\djs$ similarly to identify syntactic (non-meaningful, or~non-semantic) information similarities between planetary~spectra. 

\citet{sandford2021planetary} adopt a similar strategy, demonstrating that patterns in the information of planetary systems can reveal the mass and radius of a missing planet without input physics. Similarly, we contend that the information content of exoplanet atmospheres as contained in its modal fraction and $\djs$ can help identify exoplanets with potential habitability without the need for prior knowledge of a planet's physical properties. Indeed, with~a large enough database, the~information encoded in the spectrum will help elucidate some of these properties. (Assuming, of~course, that the spectrum has sufficient resolution).

Similar complexity measures, pattern recognition, and~information theory strategies have also been proposed for a limited selection of other astrobiology purposes. These include, amongst others, in~situ complexity measures \citep{Guttenberg2021, Marshall2021, Chou2021}, time series information content of planetary reflectance spectra \citep{Bartlett2022}, information gain as a tool for determining the observational parameters necessary to observe biosignature gases \citep{Fields_2023}, semantic (meaningful) information to describe exchange on information in daisy world models \citep{sowinski2024}, detailed chemical networks as potential biosignatures \citep{Wong2023}, and~network analysis to distinguish between biogenic and abiotic sources of atmospheric chemistry \citep{fisher2023complexsystemsapproachexoplanet}. We especially highlight the related work in \citet{guez2024}, who found that a spectral clustering algorithm can characterize simulated JWST spectra (in particular, whether they are from oxidizing or reducing atmospheres and mixing ratios for CO$_2$ and O$_2$) agnostically of the molecular features in the~spectrum. 

Finally, we note a connection between our proposed spectral $\djs$ method and methods using the Gibbs free energy of a planetary atmosphere as a biosignature \citep{Lovelock1965, Lederberg1965, cockell2009, sagan1993, krissansentotton2018, young2024}. The~Gibbs entropy of an ensemble is given by
\begin{equation}
    S_{Gibbs} = -k_B \sum_i p_i log(p_i)
\end{equation}
for $p_i$ the $i${}{th} 
 member of a particular ensemble, and $k_B$ the Boltzmann constant. The~key difference between this expression and the Shannon entropy expression (Equation (\ref{eq:djs})) is that the Gibbs entropy is a sum over molecular species $i$, while Shannon entropy in our metric is a sum over wavenumber $\nu$. In~an idealized sense, however, the~strength (e.g., absorption) of the transmission spectrum at a given wavelength is correlated to the abundance of the molecule(s) that absorb at that wavelength. This means, in~a sufficiently high-resolution and low-noise spectrum, that Shannon entropy and Gibbs entropy are correlated. The~Gibbs free energy, $G$, of~a system is
\begin{equation}
    \Delta G = \Delta H - T \Delta S_{Gibbs}
\end{equation}
for $H$ the enthalpy of the construction of a state, and $T$ the absolute temperature. This relation between $\Delta G$ and $S_{Gibbs}$ implies a further correlation between our spectral information metric and Gibbs free energy as a potential biosignature. In~contrast to the proposed methods to assess the Gibbs free energy of exoplanet atmospheres, however, our method does not require performing a retrieval to determine the abundance of a set of predetermined molecular species, $p_i$. Rather, it determines the information content from the spectrum itself. We plan to study the correlation between Gibbs free energy and our metric, especially its dependence on observational parameters, in~future~work.

In this paper, we demonstrate the efficacy of this method using simulated exoplanet data. As~a proof of concept, we use these data to show that our method recovers the results we would expect. We first show that $\djs$ is sensitive to a wide variety of planetary parameters. This indicates that our information measure can be used to identify planetary features. This analysis also shows that $\djs$ is able to differentiate between planets with Earth-like and Jupiter-like spectral characteristics with reasonable variation in planetary features. To~further validate this ability, we use simulations of observed exoplanets---including high-interest exoplanets such as habitable zone planets Proxima b and TRAPPIST-1e---to show that our method is able to distinguish planet types with realistic data. In~both of these illustrations, $\djs$ is able to identify Earth-like and Jupiter-like planets, even without a priori knowledge of what spectral features differentiate the two planet types. This is important, as~our limited understanding of the complex scenarios with the potential to host life could cause us to miss habitable planets. More detailed simulations---including those with non-equilibrium chemistry---will allow for more accurate classification with a wider class of planets. 

We also note that we have not considered here the impact of stellar flares on TRAPPIST-1e, which may have a significant impact on its potential biosignature and even whether it may host an atmosphere \citep{Herbst_2024, doyon2024temperaterockyplanetsm}. The~analysis in this work is meant to illustrate the ability of our method; a true determination of the similarity of TRAPPIST-1e to Earth will require more detailed simulations such as \citet{lin2022} or \citet{Fauchez_2019}. 

While the $\djs$ distributions comparing rocky planets to Earth, to~Jupiter, and~to a warm Jupiter clone show overlap, we argue that most of the observed rocky planets lie at the low end of the $\djs$ distribution when compared to Earth but are more spread out when compared to Jupiter and 1200 K Jupiter. The~overlap in $\djs$ can be greatly reduced by removing the four fabricated rocky planets simulations. The~exception---the one observed exoplanet with a high $\djs$ relative to Earth---is EPIC 24983012b, a~very large, hot planet. These physical parameters bring the spectrum of the exoplanet closer to a Jupiter or hot Jupiter-like planet than to Earth. The~ability of $\djs$ to identify these physical differences in fact underscores our method---EPIC 24983012b is not~Earth-like. 

The previous observations of exoplanetary spectra use narrow bandpass filters, often with low resolution. As~a result, the~information content of these spectra is very low. Even with JWST, the~shot noise limit of transmission spectroscopy limits the signal-to-noise ratio achievable without advanced spectroscopic methods proposed in near-future missions. In~the {}{Appendix \ref{sec:appendix},} 
 we show how $\djs$ varies with a simple Gaussian noise model. In~reality, noise in a transmission spectrum depends on a complex mix of instrument factors, planetary features (such as the distance between the planet and host star), and~observational conditions (such as the number of observed transits). In~future work, we hope to use spectra with simulated noise to test the dependence of our method on the signal-to-noise ratio of next generation telescope data. This can be performed for JWST using the JWST simulators \classoption{{}{JexoSim-2.0}} \citep{sarkar_jexosim_2021} or \classoption{{}{PandExo}} \citep{batalha2017}, testing the efficacy of the $\djs$ method on JWST data. We note that JWST recently confirmed the mission's ability to resolve spectral biosignatures with detections of carbon dioxide in the atmosphere of WASP-39b~\cite{aher2022}; methane in WASP-80b \citep{bell2023}; water, carbon dioxide, sulfur dioxide, and~sulfur monoxide in the gas giant WASP-39 b \citep{aher2022, rustamkulov2023early, ahrer2023early, alderson2023early, feinstein2023early}; and methane and carbon dioxide in K2-18b~\cite{madhusudhan2023, schmidt2025}. We anticipate that the method will be most helpful in future missions designed specifically for spectroscopy of potentially Earth-like planets (identified by non-spectral parameters like their mass and orbital distance) such as~HWO.

We note that, while our method is agnostic to the features that make a planet Earth-like---including life---it is not a direct measure of biosignatures. Our method assesses the similarity of an exoplanets spectrum to that of Earth. While life has a decisive impact on this similarity, a~low $\djs$ does not, in-and-of-itself, indicate life. Research is actively being conducted into true so-called ``agnostic biosignatures''. Refs. \citep{Guttenberg2021, Marshall2021, Chou2021, Wong2023, Bartlett2022} is an incomplete list. These methods may be able to identify life without assuming the form that life takes, while our method may identify Earth-like planets without assuming what features define a planet's~Earth-likeness.

\vspace{6pt} 





\authorcontributions{Conceptualization, M.G.; data curation I.D.S.; formal analysis S.V.; investigation S.V. and I.D.S.; software S.V.; supervision M.G.; validation S.V.;  writing---original draft preparation, S.V., M.G. and~I.D.S.; writing---review and editing,  S.V. and M.G.; visualization, S.V. All authors have read and agreed to the published version of the manuscript.}

\funding{This research received no external~funding.}

\institutionalreview{{}{Not applicable.} 
}

\dataavailability{The code used to generate the figures in this work is available at {}{\url{https://github.com/saracha413/space-djs} (accessed on 24 January 2022)}
.}

\acknowledgments{We thank Lisa Kaltenegger for stimulating discussions during the early stages of this work. This work was also a part of the first author's Ph.D. thesis \citep{vannah-phd}.}

\conflictsofinterest{ 
Sara Vannah was an employee of a company named Atmospheric and Environmental Research, Inc. (AER). AER is primarily a government contractor, and~its employees work on grants from Federal funding agencies to perform innovative research in the areas of atmospheric, oceanographic, and~space weather sciences. The~remaining authors declare that the research was conducted in the absence of any commercial or financial relationships that could be construed as potential conflicts of interest.} 



\newpage
\abbreviations{Abbreviations}{
The following abbreviations are used in this manuscript:\\

\noindent 
\begin{tabular}{@{}ll}
$\djs$ & Jensen--Shannon Divergence\\
$\dkl$ & Kullback--Liebler Divergence\\
HWO & Habitable Worlds Observatory\\
ESI & Earth Similarity Index
\end{tabular}
}

\appendixtitles{yes} 
\appendixstart
\appendix

\section{{}{Noise} 
 Dependence of Jensen--Shannon~Divergence}\label{sec:appendix}
An additional advantage of using $\djs$ to compare distributions is that the measure scales predictably with noise. This allows us to determine how small the noise must be to wash out a signal. Suppose, for~simplicity, that two signals have the same signal-to-noise ratio. This means the errors on their modal fractions, $p$ and $q$, are the same. We represent this as
\begin{equation}
\begin{split}
    \djs \left( p + \delta_p || q+ \delta_q \right) &= \frac{1}{2} \sum (p+\delta_p) \log\left( \frac{p+\delta_p}{\frac{1}{2}(p+q+\delta_p+\delta_q)} \right) \\
    &+ \frac{1}{2} \sum (q+\delta_q) \log\left( \frac{q+\delta_q}{\frac{1}{2}(p+q+\delta_p+\delta_q)} \right)
\end{split}
\end{equation}
for $\delta_p$ and $\delta_q$ the noise on $p$ and $q$, respectively. We may rewrite this expression so the log terms containing $\delta_p$ or $\delta_q$ take the form log($1+x$) for $x$ a multiple of $\delta_p$ and/or $\delta_q$. \mbox{This gives}
\begin{equation}
\begin{split}
    \djs \left( p + \delta_p || q+ \delta_q \right) &= \frac{1}{2} \sum \biggl[ (p+\delta_p)\left(\log(p)+\log(2)\right) \\
    &+(q+\delta_q)\left(\log(q)+\log(2)\right) - \left(p+q+\delta_p+\delta_q\right)\log(p+q) \\
   &+ \left(q+\delta_q\right)\log\left(1+\frac{\delta_q}{q} \right) + \left(p+\delta_p\right)\log\left(1+\frac{\delta_p}{p} \right) \\
   & -\left(p+q+\delta_p+\delta_q\right)\log\left(1+\frac{\delta_p + \delta_q}{p+q} \right) \biggr]
\end{split}
\end{equation}

We may then carry out the second-order Taylor expansion, $log(1+x) \approx x - \frac{1}{2}x^2$ for small x. This gives
\begin{equation}
\begin{split}
    \djs \left( p + \delta_p || q+ \delta_q \right)  &= \frac{1}{2} \sum \biggl[ (p+\delta_p)\left(\log(p)+\log(2)\right) \\
    &+ (q+\delta_q)\left(\log(q)+\log(2)\right)  - \left(p+q+\delta_p+\delta_q\right)\log(p+q) \\
    &+  \left(q+\delta_q\right)\left(\frac{\delta_q}{q} -\frac{\delta_q^2}{q^2}\right) + \left(p+\delta_p\right)\left(\frac{\delta_p}{p} -\frac{\delta_p^2}{p^2}\right) \\
    &-\left(p+q+\delta_p+\delta_q\right)\left(\frac{\delta_p + \delta_q}{p+q}-\frac{(\delta_p + \delta_q)^2}{2(p+q)^2} \right) \biggr].
\end{split}
\end{equation}

We perform the multiplications, again keeping only terms up to the second order in $\delta$ to~find
\begin{equation}
\begin{split}
    \djs \left( p + \delta_p || q+ \delta_q \right) &= \frac{1}{2} \sum \biggl[ (p+\delta_p)\left(\log(p)+\log(2)\right) \\
    &+ (q+\delta_q)\left(\log(q)+\log(2)\right) \\ 
     & -\left(p+q+\delta_p+\delta_q\right)\log(p+q) 
    + \frac{(\delta_p + \delta_q)^2}{2(p+q)} \biggr].
\end{split}
\end{equation}

We collect the terms by their order in $\delta$ to find
\begin{equation}
\begin{split}
    \djs \left( p + \delta_p || q+ \delta_q \right) &= \frac{1}{2} \sum \biggl[ p\log\left(\frac{p}{\frac{1}{2}(p+q)}\right) \\
     &+ q\log\left(\frac{q}{\frac{1}{2}(p+q)}\right)   + \delta_p\log\left(\frac{p}{\frac{1}{2}(p+q)}\right)  \\ 
     &+ \delta_q\log\left(\frac{q}{\frac{1}{2}(p+q)}\right)+ \frac{(\delta_p + \delta_q)^2}{2(p+q)} \biggr].
\end{split}
\end{equation}

The zeroth order term is simply $\djs \left( p|| q \right)$. The~change in $\djs$ with the addition of noise is then
\begin{equation}
\begin{split}
    \djs \left( p + \delta_p || q+ \delta_q \right) &= \frac{1}{2}\djs\left(p||q\right) \\ 
    &+ \frac{1}{2}\sum \left[ \delta_p\log\left(\frac{p}{\frac{1}{2}(p+q)}\right) + \delta_q\log\left(\frac{q}{\frac{1}{2}(p+q)}\right)+ \frac{(\delta_p + \delta_q)^2}{2(p+q)} \right].
    \end{split}
\end{equation}

\begin{figure}[H]

	\includegraphics[scale=0.52]{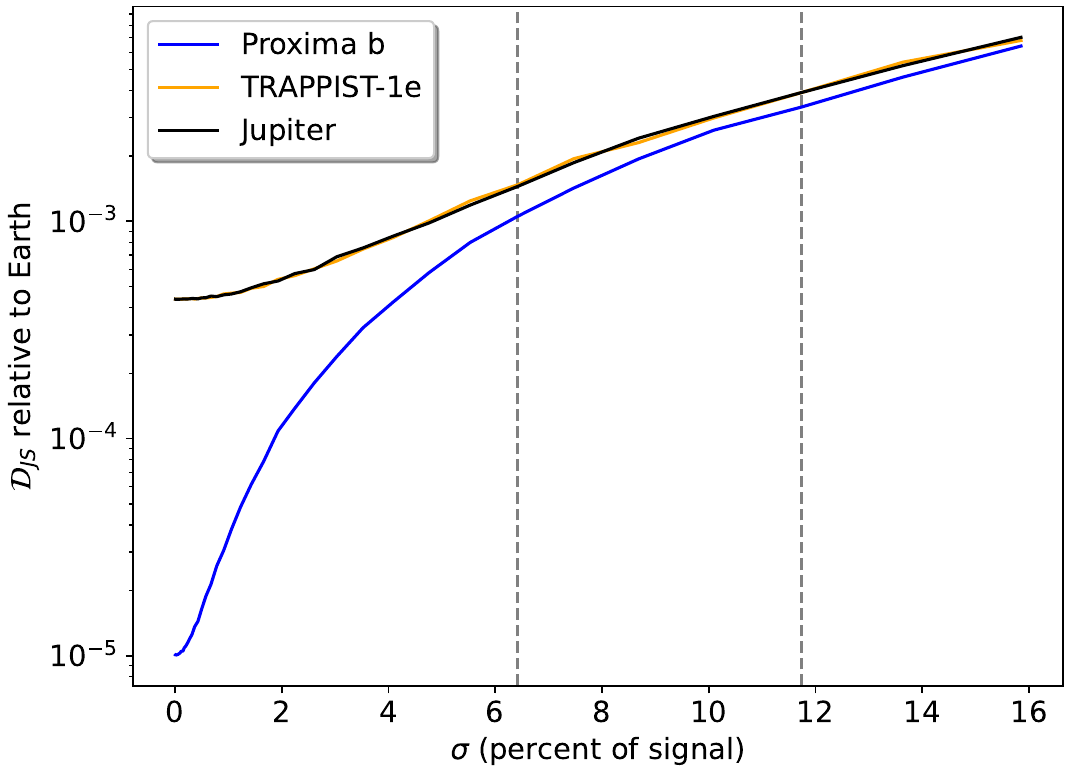}
	\caption{$\djs$ relative to Earth of two Earth-like exoplanets---Proxima b (blue line) and TRAPPIST-1e (orange line)---and Jupiter (black line) for different strengths of noise. The~noise strength is changed by choosing $\sigma$, the~width of a mean-zero Gaussian distribution from which the noise at each point in the transmission spectrum is drawn, to~be a particular percentage of strength of the transmission spectrum for each planet. For~example, $\sigma$ of 20$\%$ means that a noisy transmission spectrum for Jupiter (and each of the four planets) has been created by adding noise at each point in the spectrum where the noise is drawn from a Gaussian distribution with a width set to be 20$\%$ of the mean transit depth of the Jupiter spectrum. Gray dashed lines show the values of $\sigma$ when the two Earth-like planets are within 10$\%$ of each other (left) and when Proxima b and Jupiter are within 10$\%$ of each other. These were chosen as locations to quantify how small the noise must be in order for the planets to be~distinguishable. }
	\label{fig:noise}
\end{figure}

As an example, we examine how small the noise must be in order for Earth-like planets to be distinguished both from each other and from Jupiter. We consider the two most Earth-like exoplanets in our distribution: Proxima b, a~habitable zone planet with similar mass and radius to Earth, and~TRAPPIST-1e, a~habitable zone planet slightly smaller than Earth (suppose random noise is added to the transmission spectra for each of the planets (Earth, Proxima b, TRAPPIST-1e, and~Jupiter) all drawn from the same Gaussian distribution with width $\sigma$ chosen to be a particular percentage of the mean transit depth of the transmission spectrum of the planet. We would like to know how large $\sigma$ must be to meet two different thresholds: one for the two Earth-like exoplanets to be indistinguishable (by $\djs$) from each other and another for the two Earth-like exoplanets to be indistinguishable from Jupiter. In~Figure~\ref{fig:noise}, we show how the $\djs$ (relative to Earth) varies with $\sigma$ for each planet. For~the noiseless spectra on the far left of the plot, all three planets are easily distinguished through $\djs$. As~the intensity of the noise increases to the right of the plot, the~$\djs$ of all three planets increase but~at different rates. The~more Earth-like the planet (lowest initial $\djs$), the~more sensitive $\djs$ is to noise. This means that, with~increasing noise, the~three lines approach each other, causing the $\djs$ of the planets to be indistinguishable for large~noise.

To quantify how weak the noise must be for two planets to be indistinguishably Earth-like, we arbitrarily choose two signals to be indistinguishable if their $\djs$ values relative to Earth are within 10$\%$ of each other. The~value of $\sigma$ at this threshold for distinguishing the two Earth-like planets from each other and from Jupiter is shown in gray dashed lines. For~TRAPPIST-1e to be distinguishable from Proxima b, we find that the $\sigma$ must be less than about 6.45$\%$ of the strength of the signals. For~Proxima b to be distinguishable from Jupiter,  $\sigma$ must be less than about 11.73$\%$ of the~signals. 

We note that modeling noise for observed transmission spectra is much more complicated than the Gaussian noise method shown here as a first illustration. The~noise on a transmission spectrum depends on physical parameters such as the relative size and distance of the planet and its host star, the~number of observed transits, and~an efficient modeling of instrument noise. In~future work, we plan to use the an observation simulator such as \classoption{{}{PandExo}} \citep{batalha2017} to test the efficacy of the $\djs$ method on JWST data, with~features in Earth-like planet spectra of order 10--100 ppm \citep{beichman2018} and~especially on upcoming missions such as the~HWO.

\section{{}{Simulated} 
Spectra}\label{sec:sims}

\begin{figure}[H]

    \includegraphics[scale=0.47]{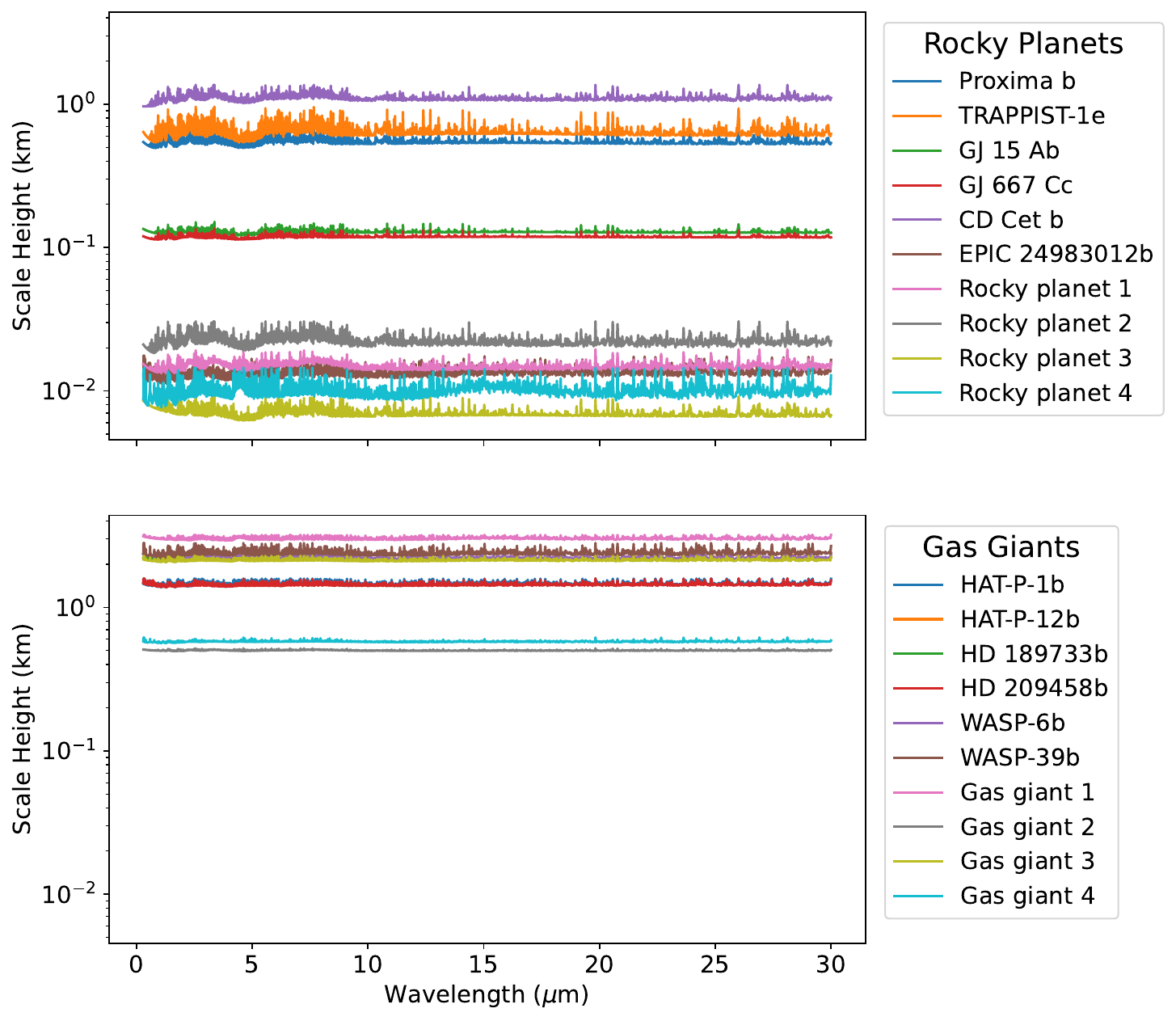}
    \caption{{Spectra} 
 for {}{rocky planets (Earth-like planets and super-Earths, top)} and gas giant planets (Jupiter-like planets and hot Jupiters, bottom) simulated by \classoption{{}{Exo-Transmit}} \citet{kempton_exo-transmit_2017}.}
    \label{fig:spectra} 
\end{figure}

The spectra simulated in this work are shown for reference in Figure~\ref{fig:spectra}.

\isPreprints{}{
\begin{adjustwidth}{-\extralength}{0cm}
} 

\reftitle{References}



\PublishersNote{}
\end{adjustwidth}

\begin{thebibliography}{999}

\bibitem[Tinetti et~al.(2018)Tinetti, Drossart, Eccleston, Hartogh, Heske,
Leconte, Micela, Ollivier, Pilbratt, Puig, et~al.]{tinetti2018chemical}
Tinetti, G.; Drossart, P.; Eccleston, P.; Hartogh, P.; Heske, A.; Leconte, J.;
Micela, G.; Ollivier, M.; Pilbratt, G.; Puig, L.;  et~al.
\newblock A chemical survey of exoplanets with ARIEL.
\newblock {\em Exp. Astron.} {\bf 2018}, {\em 46},~135--209. [\href{http://doi.org/10.1007/s10686-018-9598-x}{CrossRef}]

\bibitem[{Quanz, S. P.} et~al.(2022){Quanz, S. P.}, {Ottiger, M.}, {Fontanet,
E.}, {Kammerer, J.}, {Menti, F.}, {Dannert, F.}, {Gheorghe, A.}, {Absil, O.},
{Airapetian, V. S.}, {Alei, E.}, {Allart, R.}, {Angerhausen, D.},
{Blumenthal, S.}, {Buchhave, L. A.}, {Cabrera, J.}, {Carrión-González,
Ó.}, {Chauvin, G.}, {Danchi, W. C.}, {Dandumont, C.}, {Defrére, D.}, {Dorn,
C.}, {Ehrenreich, D.}, {Ertel, S.}, {Fridlund, M.}, {García Muñoz, A.},
{Gascón, C.}, {Girard, J. H.}, {Glauser, A.}, {Grenfell, J. L.}, {Guidi,
G.}, {Hagelberg, J.}, {Helled, R.}, {Ireland, M. J.}, {Janson, M.},
{Kopparapu, R. K.}, {Korth, J.}, {Kozakis, T.}, {Kraus, S.}, {Léger, A.},
{Leedjärv, L.}, {Lichtenberg, T.}, {Lillo-Box, J.}, {Linz, H.}, {Liseau,
R.}, {Loicq, J.}, {Mahendra, V.}, {Malbet, F.}, {Mathew, J.}, {Mennesson,
B.}, {Meyer, M. R.}, {Mishra, L.}, {Molaverdikhani, K.}, {Noack, L.}, {Oza,
A. V.}, {Pallé, E.}, {Parviainen, H.}, {Quirrenbach, A.}, {Rauer, H.},
{Ribas, I.}, {Rice, M.}, {Romagnolo, A.}, {Rugheimer, S.}, {Schwieterman, E.
W.}, {Serabyn, E.}, {Sharma, S.}, {Stassun, K. G.}, {Szulágyi, J.}, {Wang,
H. S.}, {Wunderlich, F.}, {Wyatt, M. C.}, and {the LIFE
Collaboration}]{quanz2022}
{Quanz, S.P.}; {Ottiger, M.}; {Fontanet, E.}; {Kammerer, J.}; {Menti,
F.}; {Dannert, F.}; {Gheorghe, A.}; {Absil, O.}; {Airapetian, V.S.};
\mbox{{Alei, E.};  et~al.}
\newblock Large Interferometer For Exoplanets (LIFE)---I. Improved exoplanet
detection yield estimates for a large mid-infrared space-interferometer
mission.
\newblock {\em Astron. Astrophys.} {\bf 2022}, {\em 664},~A21. [\href{http://dx.doi.org/10.1051/0004-6361/202140366}{CrossRef}]

\bibitem[National Academies~of Sciences and Medicine(2023)]{decadal}
National Academies of Sciences, Engineering, and Medicine.
\newblock {\em Pathways to Discovery in Astronomy and Astrophysics for the
2020s}; The National Academies Press: Washington, DC, USA, 2023. [\href{http://dx.doi.org/10.17226/26141}{CrossRef}]

\bibitem[Harada et~al.(2024)Harada, Dressing, Kane, and Ardestani]{Harada2024}
Harada, C.K.; Dressing, C.D.; Kane, S.R.; Ardestani, B.A.
\newblock Setting the Stage for the Search for Life with the Habitable Worlds
Observatory: Properties of 164 Promising Planet-survey Targets.
\newblock {\em  Astrophys. J. Suppl. Ser.} {\bf 2024}, {\em
272},~30. [\href{http://dx.doi.org/10.3847/1538-4365/ad3e81}{CrossRef}]

\bibitem[Ge et~al.(2022)Ge, Zhang, Zang, Deng, Mao, Xie, Liu, Zhou, Willis,
Huang, Howell, Feng, Zhu, Yao, Liu, Aizawa, Zhu, Li, Ma, Ye, Yu, Xiang, Yu,
Liu, Yang, Wang, Shi, Fang, Zong, Liu, Zhang, Zhang, El-Badry, Shen, Tam, Hu,
Yang, Zou, Wu, Lei, Wei, Wu, Sun, Wang, Zhang, Xu, Yang, Li, Xiang, Wang,
Wang, Zhang, Jia, Yuan, Zhang, Wang, Gan, Wang, Zhao, Liu, Wei, Kang, Yang,
Qi, Liu, Zhang, Zhu, Zhou, Zhang, Yu, Zhang, Li, Tang, Wang, Wang, Li, Cheng,
Shen, Li, Pan, Yang, Gao, Song, Wang, Zhang, Chen, Wang, Zhang, Wang, Zeng,
Zheng, Zhu, Guo, Zhang, Li, Wen, Feng, Chen, Chen, Han, Yang, Wang, Duan,
Huang, Liang, Bi, Gai, Ge, Guo, Huang, Li, Li, Li, Yuxi, {Lu}, Rix, Shi,
Song, Tang, Ting, Wu, Wu, Yang, Yin, Gould, Lee, Dong, Yee, Shvartzvald,
Yang, Kuang, Zhang, Liao, Qi, Yang, Zhang, Jiang, Ou, Li, Beck, Bedding,
Campante, Chaplin, Christensen-Dalsgaard, García, Gaulme, Gizon, Hekker,
Huber, Khanna, Li, Mathur, Miglio, Mosser, Ong, Santos, Stello, Bowman,
Lares-Martiz, Murphy, Niu, Ma, Molnár, Fu, De~Cat, Su, and
consortium]{ye2022china}
Ge, J.; Zhang, H.; Zang, W.; Deng, H.; Mao, S.; Xie, J.W.; Liu, H.G.; Zhou,
J.L.; Willis, K.; Huang, C.;  et~al.
\newblock ET White Paper: To Find the First Earth 2.0. \emph{arXiv} {\bf 2022}, arXiv:2206.06693. [\href{http://dx.doi.org/10.48550/arxiv.2206.06693}{CrossRef}]

\bibitem[Ramsay et~al.(2020)Ramsay, Amico, Bezawada, Cirasuolo, Derie, Egner,
George, Gont{\'e}, Herrera, Hammersley, et~al.]{ramsay2020eso}
Ramsay, S.; Amico, P.; Bezawada, N.; Cirasuolo, M.; Derie, F.; Egner, S.;
George, E.; Gont{\'e}, F.; Herrera, J.C.G.; \mbox{Hammersley, P.;  et~al.}
\newblock The ESO extremely large telescope instrumentation programme.
\newblock In Proceedings of the Advances in Optical Astronomical
Instrumentation 2019, {}{Melbourne, Australia,  8--12 December 2019}; International Society for Optics and Photonics: {}{Bellingham, WA, USA},  2020;
Volume 11203, p. 1120303. [\href{http://dx.doi.org/10.1117/12.2541400}{CrossRef}]

\bibitem[Rustamkulov et~al.(2023)Rustamkulov, Sing, Mukherjee, May, Kirk,
Schlawin, Line, Piaulet, Carter, Batalha, Goyal, L{\'o}pez-Morales,
Lothringer, MacDonald, Moran, Stevenson, Wakeford, Espinoza, Bean, Batalha,
Benneke, Berta-Thompson, Crossfield, Gao, Kreidberg, Powell, Cubillos,
Gibson, Leconte, Molaverdikhani, Nikolov, Parmentier, Roy, Taylor, Turner,
Wheatley, Aggarwal, Ahrer, Alam, Alderson, Allen, Banerjee, Barat, Barrado,
Barstow, Bell, Blecic, Brande, Casewell, Changeat, Chubb, Crouzet, Daylan,
Decin, D{\'e}sert, Mikal-Evans, Feinstein, Flagg, Fortney, Harrington, Heng,
Hong, Hu, Iro, Kataria, Kempton, Krick, Lendl, Lillo-Box, Louca,
Lustig-Yaeger, Mancini, Mansfield, Mayne, Miguel, Morello, Ohno, Palle, Petit
dit de~la Roche, Rackham, Radica, Ramos-Rosado, Redfield, Rogers, Shkolnik,
Southworth, Teske, Tremblin, Tucker, Venot, Waalkes, Welbanks, Zhang, and
Zieba]{rustamkulov2023early}
Rustamkulov, Z.; Sing, D.K.; Mukherjee, S.; May, E.M.; Kirk, J.; Schlawin, E.;
Line, M.R.; Piaulet, C.; Carter, A.L.; \mbox{Batalha, N.E.;  et~al.}
\newblock Early Release Science of the exoplanet WASP-39b with JWST NIRSpec
PRISM.
\newblock {\em Nature} {\bf 2023}, {\em 614},~659--663. [\href{http://dx.doi.org/10.1038/s41586-022-05677-y}{CrossRef}]

\bibitem[Ahrer et~al.(2023)Ahrer, Stevenson, Mansfield, Moran, Brande, Morello,
Murray, Nikolov, Petit Dit de~la Roche, Schlawin, et~al.]{ahrer2023early}
Ahrer, E.M.; Stevenson, K.B.; Mansfield, M.; Moran, S.E.; Brande, J.; Morello,
G.; Murray, C.A.; Nikolov, N.K.; Petit Dit de~la Roche, D.J.; Schlawin, E.;
et~al.
\newblock Early Release Science of the exoplanet WASP-39b with JWST NIRCam.
\newblock {\em Nature} {\bf 2023}, \emph{614}, 653--658. [\href{http://dx.doi.org/10.1038/s41586-022-05590-4}{CrossRef}]

\bibitem[Alderson et~al.(2023)Alderson, Wakeford, Alam, Batalha, Lothringer,
Adams~Redai, Barat, Brande, Damiano, Daylan, et~al.]{alderson2023early}
Alderson, L.; Wakeford, H.R.; Alam, M.K.; Batalha, N.E.; Lothringer, J.D.;
Adams~Redai, J.; Barat, S.; Brande, J.; Damiano, M.; Daylan, T.;  et~al.
\newblock Early Release Science of the exoplanet WASP-39b with JWST NIRSpec
G395H.
\newblock {\em Nature} {\bf 2023}, {\em 614},~664--669. [\href{http://dx.doi.org/10.1038/s41586-022-05591-3}{CrossRef}]

\bibitem[Feinstein et~al.(2023)Feinstein, Radica, Welbanks, Murray, Ohno,
Coulombe, Espinoza, Bean, Teske, Benneke, et~al.]{feinstein2023early}
Feinstein, A.D.; Radica, M.; Welbanks, L.; Murray, C.A.; Ohno, K.; Coulombe,
L.P.; Espinoza, N.; Bean, J.L.; Teske, J.K.; \mbox{Benneke, B.;  et~al.}
\newblock Early Release Science of the exoplanet WASP-39b with JWST NIRISS.
\newblock {\em Nature} {\bf 2023}, {\em 614},~670--675. [\href{http://dx.doi.org/10.1038/s41586-022-05674-1}{CrossRef}]

\bibitem[Pontoppidan et~al.(2022)Pontoppidan, Barrientes, Blome, Braun, Brown,
Carruthers, Coe, DePasquale, Espinoza, Marin, Gordon, Henry, Hustak, James,
Jenkins, Koekemoer, LaMassa, Law, Lockwood, Moro-Martin, Mullally, Pagan,
Player, Proffitt, Pulliam, Ramsay, Ravindranath, Reid, Robberto, Sabbi,
Ubeda, Balogh, Flanagan, Gardner, Hasan, Meinke, and Nota]{Pontoppidan_2022}
Pontoppidan, K.M.; Barrientes, J.; Blome, C.; Braun, H.; Brown, M.; Carruthers,
M.; Coe, D.; DePasquale, J.; Espinoza, N.; \mbox{Marin, M.G.;  et~al.}
\newblock The JWST Early Release Observations.
\newblock {\em  Astrophys. J. Lett.} {\bf 2022}, {\em 936},~L14. [\href{http://dx.doi.org/10.3847/2041-8213/ac8a4e}{CrossRef}]

\bibitem[Madhusudhan et~al.(2020)Madhusudhan, Nixon, Welbanks, Piette, and
Booth]{Madhusudhan_2020}
Madhusudhan, N.; Nixon, M.C.; Welbanks, L.; Piette, A.A.A.; Booth, R.A.
\newblock The Interior and Atmosphere of the Habitable-zone Exoplanet K2-18b.
\newblock {\em  Astrophys. J. Lett.} {\bf 2020}, {\em 891},~L7. [\href{http://dx.doi.org/10.3847/2041-8213/ab7229}{CrossRef}]

\bibitem[Schulze-Makuch et~al.(2011)Schulze-Makuch, M\'{e}ndez, Fair\'{e}n, von
Paris, Turse, Boyer, Davila, Ant\'{o}nio, Catling, and
Irwin]{schulzemakuch2011}
Schulze-Makuch, D.; M\'{e}ndez, A.; Fair\'{e}n, A.G.; von Paris, P.; Turse, C.;
Boyer, G.; Davila, A.F.; Ant\'{o}nio, M.R.d.S.; Catling, D.; Irwin, L.N.
\newblock A Two-Tiered Approach to Assessing the Habitability of Exoplanets.
\newblock {\em Astrobiology} {\bf 2011}, {\em 11},~1041--1052. [\href{http://dx.doi.org/10.1089/ast.2010.0592}{CrossRef}] [\href{http://www.ncbi.nlm.nih.gov/pubmed/22017274}{PubMed}]

\bibitem[Kashyap~Jagadeesh et~al.(2017)Kashyap~Jagadeesh, Gudennavar, Doshi,
and Safonova]{Kashyap2017}
Kashyap~Jagadeesh, M.; Gudennavar, S.B.; Doshi, U.; Safonova, M.
\newblock Indexing of exoplanets in search for potential habitability:
Application to Mars-like worlds.
\newblock {\em Astrophys. Space Sci.} {\bf 2017}, {\em 362},~146. [\href{http://dx.doi.org/10.1007/s10509-017-3131-y}{CrossRef}]

\bibitem[Jagadeesh(2018)]{jagadeesh2018earthsimilarityindexhabitability}
Jagadeesh, M.K.
\newblock Earth Similarity Index and Habitability Studies of Exoplanets. \emph{arXiv} \textbf{2018}, 	{}{arXiv:1801.07101}. [\href{http://dx.doi.org/10.48550/arXiv.1801.07101}{CrossRef}]

\bibitem[Basak et~al.(2020)Basak, Saha, Mathur, Bora, Makhija, Safonova, and
Agrawal]{basak2020}
Basak, S.; Saha, S.; Mathur, A.; Bora, K.; Makhija, S.; Safonova, M.; Agrawal,
S.
\newblock CEESA meets machine learning: A Constant Elasticity Earth Similarity
Approach to habitability and classification of exoplanets.
\newblock {\em Astron. Comput.} {\bf 2020}, {\em 30},~100335. [\href{http://dx.doi.org/10.1016/j.ascom.2019.100335}{CrossRef}]

\bibitem[Bora et~al.(2016)Bora, Saha, Agrawal, Safonova, Routh, and
Narasimhamurthy]{bora2016}
Bora, K.; Saha, S.; Agrawal, S.; Safonova, M.; Routh, S.; Narasimhamurthy, A.
\newblock CD-HPF: New habitability score via data analytic modeling.
\newblock {\em Astron. Comput.} {\bf 2016}, {\em 17},~129--143. [\href{http://dx.doi.org/10.1016/j.ascom.2016.08.001}{CrossRef}]

\bibitem[Irwin et~al.(2014)Irwin, Méndez, Fairén, and
Schulze-Makuch]{challe5010159}
Irwin, L.N.; Méndez, A.; Fairén, A.G.; Schulze-Makuch, D.
\newblock Assessing the Possibility of Biological Complexity on Other Worlds,
with an Estimate of the Occurrence of Complex Life in the Milky Way Galaxy.
\newblock {\em Challenges} {\bf 2014}, {\em 5},~159--174. [\href{http://dx.doi.org/10.3390/challe5010159}{CrossRef}]

\bibitem[Sarkar et~al.(2021)Sarkar, Bhatia, Saha, Safonova, and
Sarkar]{Sarkar2021}
Sarkar, J.; Bhatia, K.; Saha, S.; Safonova, M.; Sarkar, S.
\newblock Postulating exoplanetary habitability via a novel anomaly detection
method.
\newblock {\em Mon. Not. R. Astron. Soc.} {\bf 2021},
{\em 510},~6022–6032. [\href{http://dx.doi.org/10.1093/mnras/stab3556}{CrossRef}]

\bibitem[{Saha} et~al.(2018){Saha}, {Basak}, {Safonova}, {Bora}, {Agrawal},
{Sarkar}, and {Murthy}]{saha2018}
{Saha}, S.; {Basak}, S.; {Safonova}, M.; {Bora}, K.; {Agrawal}, S.; {Sarkar},
P.; {Murthy}, J.
\newblock {Theoretical validation of potential habitability via analytical and
boosted tree methods: An optimistic study on recently discovered exoplanets}.
\newblock {\em Astron. Comput.} {\bf 2018}, {\em 23},~141. [\href{http://dx.doi.org/10.1016/j.ascom.2018.03.003}{CrossRef}]

\bibitem[Seager(2014)]{seager2014future}
Seager, S.
\newblock The future of spectroscopic life detection on exoplanets.
\newblock {\em Proc. Natl. Acad. Sci. USA} {\bf 2014},
{\em 111},~12634--12640. [\href{http://dx.doi.org/10.1073/pnas.1304213111}{CrossRef}]

\bibitem[Seager(2013)]{seager2013exoplanet}
Seager, S.
\newblock Exoplanet habitability.
\newblock {\em Science} {\bf 2013}, {\em 340},~577--581. [\href{http://dx.doi.org/10.1126/science.1232226}{CrossRef}]

\bibitem[Schwieterman et~al.(2018)Schwieterman, Kiang, Parenteau, Harman,
DasSarma, Fisher, Arney, Hartnett, Reinhard, Olson, et~al.]{Schwieterman2018}
Schwieterman, E.W.; Kiang, N.Y.; Parenteau, M.N.; Harman, C.E.; DasSarma, S.;
Fisher, T.M.; Arney, G.N.; Hartnett, H.E.; Reinhard, C.T.; Olson, S.L.;
et~al.
\newblock Exoplanet biosignatures: A review of remotely detectable signs of
life.
\newblock {\em Astrobiology} {\bf 2018}, {\em 18},~663--708. [\href{http://dx.doi.org/10.1089/ast.2017.1729}{CrossRef}] [\href{http://www.ncbi.nlm.nih.gov/pubmed/29727196}{PubMed}]

\bibitem[Seager et~al.(2005)Seager, Turner, Schafer, and Ford]{seager2005}
Seager, S.; Turner, E.; Schafer, J.; Ford, E.
\newblock Vegetation's Red Edge: A Possible Spectroscopic Biosignature of
Extraterrestrial Plants.
\newblock {\em Astrobiology} {\bf 2005}, {\em 5},~372--390. [\href{http://dx.doi.org/10.1089/ast.2005.5.372}{CrossRef}] [\href{http://www.ncbi.nlm.nih.gov/pubmed/15941381}{PubMed}]

\bibitem[O'Malley-James and Kaltenegger(2018)]{omalley2018}
O'Malley-James, J.T.; Kaltenegger, L.
\newblock The Vegetation Red Edge Biosignature Through Time on Earth and
Exoplanets.
\newblock {\em Astrobiology} {\bf 2018}, {\em 18},~1123--1136. [\href{http://dx.doi.org/10.1089/ast.2017.1798}{CrossRef}] [\href{http://www.ncbi.nlm.nih.gov/pubmed/30204495}{PubMed}]

\bibitem[Sagan et~al.(1993)Sagan, Thompson, Carlson, Gurnett, and
Hord]{sagan1993}
Sagan, C.; Thompson, W.R.; Carlson, R.; Gurnett, D.; Hord, C.
\newblock A search for life on Earth from the Galileo spacecraft.
\newblock {\em Nature} {\bf 1993}, {\em 365},~715--721. [\href{http://dx.doi.org/10.1038/365715a0}{CrossRef}]

\bibitem[Thompson et~al.(2022)Thompson, Krissansen-Totton, Wogan, Telus, and
Fortney]{Thompson2022}
Thompson, M.A.; Krissansen-Totton, J.; Wogan, N.; Telus, M.; Fortney, J.J.
\newblock The case and context for atmospheric methane as an exoplanet
biosignature.
\newblock {\em Proc. Natl. Acad. Sci. USA} {\bf 2022},
{\em 119},~e2117933119. [\href{http://dx.doi.org/10.1073/pnas.2117933119}{CrossRef}]

\bibitem[Tinetti et~al.(2021)Tinetti, Eccleston, Haswell, Lagage, Leconte,
Lüftinger, Micela, Min, Pilbratt, Puig, Swain, Testi, Turrini,
Vandenbussche, Osorio, Aret, Beaulieu, Buchhave, Ferus, Griffin, Guedel,
Hartogh, Machado, Malaguti, Pallé, Rataj, Ray, Ribas, Szabó, Tan, Werner,
Ratti, Scharmberg, Salvignol, Boudin, Halain, Haag, Crouzet, Kohley, Symonds,
Renk, Caldwell, Abreu, Alonso, Amiaux, Berthé, Bishop, Bowles, Carmona,
Coffey, Colomé, Crook, Désjonqueres, Díaz, Drummond, Focardi, Gómez,
Holmes, Krijger, Kovacs, Hunt, Machado, Morgante, Ollivier, Ottensamer, Pace,
Pagano, Pascale, Pearson, Pedersen, Pniel, Roose, Savini, Stamper,
Szirovicza, Szoke, Tosh, Vilardell, Barstow, Borsato, Casewell, Changeat,
Charnay, Civiš, du~Foresto, Coustenis, Cowan, Danielski, Demangeon,
Drossart, Edwards, Gilli, Encrenaz, Kiss, Kokori, Ikoma, Morales, Mendonça,
Moneti, Mugnai, Muñoz, Helled, Kama, Miguel, Nikolaou, Pagano, Panic,
Rengel, Rickman, Rocchetto, Sarkar, Selsis, Tennyson, Tsiaras, Venot, Vida,
Waldmann, Yurchenko, Szabó, Zellem, Al-Refaie, Alvarez, Anisman, Arhancet,
Ateca, Baeyens, Barnes, Bell, Benatti, Biazzo, Błęcka, Bonomo, Bosch,
Bossini, Bourgalais, Brienza, Brucalassi, Bruno, Caines, Calcutt, Campante,
Canestrari, Cann, Casali, Casas, Cassone, Cara, Carmona, Carone, Carrasco,
Changeat, Chioetto, Cortecchia, Czupalla, Chubb, Ciaravella, Claret, Claudi,
Codella, Comas, Cracchiolo, Cubillos, Peppo, Decin, Dejabrun, Delgado-Mena,
Giorgio, Diolaiti, Dorn, Doublier, Doumayrou, Dransfield, Dumaye, Dunford,
Escobar, Eylen, Farina, Fedele, Fernández, Fleury, Fonte, Fontignie,
Fossati, Funke, Galy, Garai, García, García-Rigo, Garufi, Sacco, Giacobbe,
Gómez, Gonzalez, Gonzalez-Galindo, Grassi, Griffith, Guarcello, Goujon,
Gressier, Grzegorczyk, Guillot, Guilluy, Hargrave, Hellin, Herrero, Hills,
Horeau, Ito, Jessen, Kabath, Kálmán, Kawashima, Kimura, Knížek,
Kreidberg, Kruid, Kruijssen, Kubelík, Lara, Lebonnois, Lee, Lefevre,
Lichtenberg, Locci, Lombini, Lopez, Lorenzani, MacDonald, Magrini, Maldonado,
Marcq, Migliorini, Modirrousta-Galian, Molaverdikhani, Molinari, Mollière,
Moreau, Morello, Morinaud, Morvan, Moses, Mouzali, Nakhjiri, Naponiello,
Narita, Nascimbeni, Nikolaou, Noce, Oliva, Palladino, Papageorgiou,
Parmentier, Peres, Pérez, Perez-Hoyos, Perger, Pestellini, Petralia,
Philippon, Piccialli, Pignatari, Piotto, Podio, Polenta, Preti, Pribulla,
Puertas, Rainer, Reess, Rimmer, Robert, Rosich, Rossi, Rust, Saleh, Sanna,
Schisano, Schreiber, Schwartz, Scippa, Seli, Shibata, Simpson, Shorttle,
Skaf, Skup, Sobiecki, Sousa, Sozzetti, Šponer, Steiger, Tanga, Tackley,
Taylor, Tecza, Terenzi, Tremblin, Tozzi, Triaud, Trompet, Tsai, Tsantaki,
Valencia, Vandaele, der Swaelmen, Vardan, Vasisht, Vazan, Vecchio, Waltham,
Wawer, Widemann, Wolkenberg, Yip, Yung, Zilinskas, Zingales, and
Zuppella]{Tinetti2021}
Tinetti, G.; Eccleston, P.; Haswell, C.; Lagage, P.O.; Leconte, J.; Lüftinger,
T.; Micela, G.; Min, M.; Pilbratt, G.; Puig, L.;  et~al.
\newblock Ariel: Enabling planetary science across light-years. \emph{arXiv} {\bf 2021}, arXiv:2104.04824. [\href{http://dx.doi.org/10.48550/arXiv.2104.04824}{CrossRef}]

\bibitem[Udry et~al.(2014)Udry, Lovis, Bouchy, Cameron, Henning, Mayor, Pepe,
Piskunov, Pollacco, Queloz, et~al.]{udry2014exoplanet}
Udry, S.; Lovis, C.; Bouchy, F.; Cameron, A.C.; Henning, T.; Mayor, M.; Pepe,
F.; Piskunov, N.; Pollacco, D.; Queloz, D.;  et~al.
\newblock Exoplanet science with the European Extremely Large Telescope. The
case for visible and near-IR spectroscopy at high resolution.
\newblock {\em arXiv} {\bf 2014}, arXiv:1412.1048. [\href{http://dx.doi.org/10.17863/CAM.8917}{CrossRef}]

\bibitem[{Wang} et~al.(2018){Wang}, {Mawet}, {Hu}, {Ruane}, {Delorme}, and
{Klimovich}]{wang2018}
{Wang}, J.; {Mawet}, D.; {Hu}, R.; {Ruane}, G.; {Delorme}, J.R.; {Klimovich},
N.
\newblock {Baseline requirements for detecting biosignatures with the HabEx and
LUVOIR mission concepts}.
\newblock {\em J. Astron. Telesc. Instrum. Syst.}
{\bf 2018}, {\em 4},~035001. [\href{http://dx.doi.org/10.1117/1.JATIS.4.3.035001}{CrossRef}]

\bibitem[Vannah et~al.(2023)Vannah, Gleiser, and Kaltenegger]{vannah2024}
Vannah, S.; Gleiser, M.; Kaltenegger, L.
\newblock An information theory approach to identifying signs of life on
transiting planets.
\newblock {\em Mon. Not. R. Astron. Soc. Lett.} {\bf
2023}, {\em 528},~L4--L9. [\href{http://dx.doi.org/10.1093/mnrasl/slad156}{CrossRef}]

\bibitem[Gleiser and Stamatopoulos(2012)]{Gleiser2012}
Gleiser, M.; Stamatopoulos, N.
\newblock Entropic Measure for Localized Energy Configurations: Kinks, Bounces,
and Bubbles.
\newblock \mbox{{\em Phys. Lett. B}} {\bf 2012}, {\em 713},~304--307. [\href{http://dx.doi.org/10.1016/j.physletb.2012.05.064}{CrossRef}]

\bibitem[Shannon(1948)]{Shannon1948}
Shannon, C.E.
\newblock A Mathematical Theory of Communication.
\newblock {\em Bell Syst. Tech. J.} {\bf 1948}, {\em 27},~623--656. [\href{http://dx.doi.org/10.1002/j.1538-7305.1948.tb00917.x}{CrossRef}]

\bibitem[Lin(1991)]{lin1991divergence}
Lin, J.
\newblock Divergence measures based on the Shannon entropy.
\newblock {\em IEEE Trans. Inf. Theory} {\bf 1991}, {\em
37},~145--151. [\href{http://dx.doi.org/10.1109/18.61115}{CrossRef}]

\bibitem[Kullback and Leibler(1951)]{Kullback:1951}
Kullback, S.; Leibler, R.A.
\newblock {On Information and Sufficiency}.
\newblock {\em  Ann. Math. Stat.} {\bf 1951}, {\em
22},~79--86. [\href{http://dx.doi.org/10.1214/aoms/1177729694}{CrossRef}]

\bibitem[Gleiser and Stamatopoulos(2012)]{Gleiser2012a}
Gleiser, M.; Stamatopoulos, N.
\newblock Information content of spontaneous symmetry breaking.
\newblock {\em Phys. Rev. D} {\bf 2012}, {\em 86},~045004. [\href{http://dx.doi.org/10.1103/PhysRevD.86.045004}{CrossRef}]

\bibitem[Kempton et~al.(2017)Kempton, Lupu, Owusu-Asare, Slough, and
Cale]{kempton_exo-transmit_2017}
Kempton, E.M.R.; Lupu, R.; Owusu-Asare, A.; Slough, P.; Cale, B.
\newblock Exo-{Transmit}: {An} {Open}-{Source} {Code} for {Calculating}
{Transmission} {Spectra} for {Exoplanet} {Atmospheres} of {Varied}
{Composition}.
\newblock {\em Publ. Astron. Soc. Pac.} {\bf
2017}, {\em 129},~044402. [\href{http://dx.doi.org/10.1088/1538-3873/aa61ef}{CrossRef}]

\bibitem[Lustig-Yaeger et~al.(2023)Lustig-Yaeger, Meadows, Crisp, Line, and
Robinson]{Lustig-Yaeger_2023}
Lustig-Yaeger, J.; Meadows, V.S.; Crisp, D.; Line, M.R.; Robinson, T.D.
\newblock Earth as a Transiting Exoplanet: A Validation of Transmission
Spectroscopy and Atmospheric Retrieval Methodologies for Terrestrial
Exoplanets.
\newblock {\em  Planet. Sci. J.} {\bf 2023}, {\em 4},~170. [\href{http://dx.doi.org/10.3847/PSJ/acf3e5}{CrossRef}]

\bibitem[Macdonald and Cowan(2019)]{macdonald2019}
Macdonald, E.J.R.; Cowan, N.B.
\newblock An empirical infrared transit spectrum of Earth: Opacity windows and
biosignatures.
\newblock {\em Mon. Not. R. Astron. Soc.} {\bf 2019},
{\em 489},~196--204. [\href{http://dx.doi.org/10.1093/mnras/stz2047}{CrossRef}]

\bibitem[Kaltenegger and Traub(2009)]{Kaltenegger_2009}
Kaltenegger, L.; Traub, W.A.
\newblock Transits of Earth-Like Planets.
\newblock {\em  Astrophys. J.} {\bf 2009}, {\em 698},~519. [\href{http://dx.doi.org/10.1088/0004-637X/698/1/519}{CrossRef}]

\bibitem[Robinson et~al.(2011)Robinson, Meadows, Crisp, Deming, A'Hearn,
Charbonneau, Livengood, Seager, Barry, Hearty, Hewagama, Lisse, McFadden, and
Wellnitz]{robinson2011}
Robinson, T.D.; Meadows, V.S.; Crisp, D.; Deming, D.; A'Hearn, M.F.;
Charbonneau, D.; Livengood, T.A.; Seager, S.; Barry, R.K.; Hearty, T.;
et~al.
\newblock Earth as an Extrasolar Planet: Earth Model Validation Using EPOXI
Earth Observations.
\newblock {\em Astrobiology} {\bf 2011}, {\em 11},~393--408. [\href{http://dx.doi.org/10.1089/ast.2011.0642}{CrossRef}] [\href{http://www.ncbi.nlm.nih.gov/pubmed/21631250}{PubMed}]

\bibitem[Kaltenegger et~al.(2007)Kaltenegger, Traub, and
Jucks]{Kaltenegger_2007}
Kaltenegger, L.; Traub, W.A.; Jucks, K.W.
\newblock Spectral Evolution of an Earth-like Planet.
\newblock {\em  Astrophys. J.} {\bf 2007}, {\em 658},~598. [\href{http://dx.doi.org/10.1086/510996}{CrossRef}]

\bibitem[Montañes-Rodriguez et~al.(2015)Montañes-Rodriguez, Gonzalez-Merino,
Palle, Lopez-Puertas, and Garcia-Melendo]{montanes-rodriguez_jupiter_2015}
Montañes-Rodriguez, P.; Gonzalez-Merino, B.; Palle, E.; Lopez-Puertas, M.;
Garcia-Melendo, E.
\newblock Jupiter as an exoplanet: {UV} to {NIR} transmission spectrum reveals
hazes, a {Na} layer and possibly stratospheric {H\textsubscript{2}O}-ice clouds.
\newblock {\em  Astrophys. J.} {\bf 2015}, {\em 801},~L8. [\href{http://dx.doi.org/10.1088/2041-8205/801/1/L8}{CrossRef}]

\bibitem[Sato and Hansen(1979)]{sato_jupiters_1979}
Sato, M.; Hansen, J.E.
\newblock Jupiter's {Atmospheric} {Composition} and {Cloud} {Structure}
{Deduced} from {Absorption} {Bands} in {Reflected} {Sunlight}.
\newblock {\em J. Atmos. Sci.} {\bf 1979}, {\em
36},~1133--1167. [\href{http://dx.doi.org/10.1175/1520-0469(1979)036&lt;1133:JACACS&gt;2.0.CO;2}{CrossRef}]

\bibitem[Liu et~al.(2014)Liu, Asplund, Ramirez, Yong, and
Melendez]{liu_high_2014}
Liu, F.; Asplund, M.; Ramirez, I.; Yong, D.; Melendez, J.
\newblock A high precision chemical abundance analysis of the {HAT}-{P}-1
stellar binary: Constraints on planet formation.
\newblock {\em Mon. Not. R. Astron. Soc. Lett.} {\bf
2014}, {\em 442},~L51--L55. [\href{http://dx.doi.org/10.1093/mnrasl/slu055}{CrossRef}]

\bibitem[Hartman et~al.(2009)Hartman, Bakos, Torres, Kovács, Noyes, Pál,
Latham, Sip{\textbackslash}Hocz, Fischer, Johnson, Marcy, Butler, Howard,
Esquerdo, Sasselov, Kovács, Stefanik, Fernandez, Lázár, Papp, and
Sári]{hartman_hat-p-12b_2009}
Hartman, J.D.; Bakos, G.A.; Torres, G.; Kovács, G.; Noyes, R.W.; Pál, A.;
Latham, D.W.; Sip{\textbackslash}Hocz, B.; Fischer, D.A.; \mbox{Johnson, J.A.;
et~al.}
\newblock HAT-P-12b: A low-density sub-saturn mass planet transiting a
metal-poor K dwarf.
\newblock {\em  Astrophys. J.} {\bf 2009}, {\em 706},~785--796. [\href{http://dx.doi.org/10.1088/0004-637X/706/1/785}{CrossRef}]

\bibitem[Boyajian et~al.(2015)Boyajian, von Braun, Feiden, Huber, Basu,
Demarque, Fischer, Schaefer, Mann, White, Maestro, Brewer, Lamell, Spada,
López-Morales, Ireland, Farrington, van Belle, Kane, Jones, ten Brummelaar,
Ciardi, McAlister, Ridgway, Goldfinger, Turner, and
Sturmann]{boyajian_stellar_2015}
Boyajian, T.; von Braun, K.; Feiden, G.A.; Huber, D.; Basu, S.; Demarque, P.;
Fischer, D.A.; Schaefer, G.; Mann, A.W.; \mbox{White, T.R.;  et~al.}
\newblock Stellar diameters and temperatures---{VI}. {High} angular
resolution measurements of the transiting exoplanet host stars {HD} 189733
and {HD} 209458 and implications for models of cool dwarfs.
\newblock {\em Mon. Not. R. Astron. Soc.} {\bf 2015},
{\em 447},~846--857. [\href{http://dx.doi.org/10.1093/mnras/stu2502}{CrossRef}]

\bibitem[del Burgo and Allende~Prieto(2016)]{del_burgo_accurate_2016}
Del Burgo, C.; Allende~Prieto, C.
\newblock Accurate parameters for {HD} 209458 and its planet from {HST}
spectrophotometry.
\newblock {\em Mon. Not. R. Astron. Soc.} {\bf 2016},
{\em 463},~1400--1408. [\href{http://dx.doi.org/10.1093/mnras/stw2005}{CrossRef}]

\bibitem[Gillon et~al.(2009)Gillon, Anderson, Triaud, Hellier, Maxted, Pollaco,
Queloz, Smalley, West, Wilson, Bentley, Collier~Cameron, Enoch, Hebb, Horne,
Irwin, Joshi, Lister, Mayor, Pepe, Parley, Segransan, Udry, and
Wheatley]{gillon_discovery_2009}
Gillon, M.; Anderson, D.R.; Triaud, A.H.M.J.; Hellier, C.; Maxted, P.F.L.;
Pollaco, D.; Queloz, D.; Smalley, B.; West, R.G.; \mbox{Wilson, D.M.;  et~al.}
\newblock Discovery and characterization of {WASP}-6b, an inflated
sub-{Jupiter} mass planet transiting a solar-type star.
\newblock {\em Astron. Astrophys.} {\bf 2009}, {\em 501},~785--792. [\href{http://dx.doi.org/10.1051/0004-6361/200911749}{CrossRef}]

\bibitem[Faedi et~al.(2011)Faedi, Barros, Anderson, Brown, Cameron, Pollacco,
Boisse, Hébrard, Lendl, Lister, Smalley, Street, Triaud, Bento, Bouchy,
Butters, Enoch, Haswell, Hellier, Keenan, Miller, Moulds, Moutou, Norton,
Queloz, Santerne, Simpson, Skillen, Smith, Udry, Watson, West, and
Wheatley]{faedi_wasp-39b_2011}
Faedi, F.; Barros, S.C.C.; Anderson, D.R.; Brown, D.J.A.; Cameron, A.C.;
Pollacco, D.; Boisse, I.; Hébrard, G.; Lendl, M.; \mbox{Lister, T.A.;  et~al.}
\newblock {WASP}-39b: A highly inflated {Saturn}-mass planet orbiting a late
{G}-type star.
\newblock {\em Astron. Astrophys.} {\bf 2011}, {\em 531},~A40. [\href{http://dx.doi.org/10.1051/0004-6361/201116671}{CrossRef}]

\bibitem[Lin and Kaltenegger(2020)]{lin_high-resolution_2020}
Lin, Z.; Kaltenegger, L.
\newblock High-resolution reflection spectra for {Proxima} b and {Trappist}-1e
models for {ELT} observations.
\newblock {\em Mon. Not. R. Astron. Soc.} {\bf 2020},
{\em 491},~2845--2854. [\href{http://dx.doi.org/10.1093/mnras/stz3213}{CrossRef}]

\bibitem[Barnes et~al.(2018)Barnes, Deitrick, Luger, Driscoll, Quinn, Fleming,
Guyer, McDonald, Meadows, Arney, Crisp, Domagal-Goldman, Foreman-Mackey,
Kaib, Lincowski, Lustig-Yaeger, and Schwieterman]{barnes_habitability_2016}
Barnes, R.; Deitrick, R.; Luger, R.; Driscoll, P.E.; Quinn, T.R.; Fleming,
D.P.; Guyer, B.; McDonald, D.V.; Meadows, V.S.; \mbox{Arney, G.;  et~al.}
\newblock The Habitability of Proxima Centauri b I: Evolutionary Scenarios.
\newblock \emph{arXiv}  \textbf{2018},  arXiv:1608.06919v2. [\href{http://dx.doi.org/10.48550/arXiv.1608.06919}{CrossRef}]

\bibitem[Delrez et~al.(2018)Delrez, Gillon, Triaud, Demory, de~Wit, Ingalls,
Agol, Bolmont, Burdanov, Burgasser, Carey, Jehin, Leconte, Lederer, Queloz,
Selsis, and Van~Grootel]{delrez_early_2018}
Delrez, L.; Gillon, M.; Triaud, A.H.M.J.; Demory, B.O.; de~Wit, J.; Ingalls,
J.G.; Agol, E.; Bolmont, E.; Burdanov, A.; \mbox{Burgasser, A.J.;  et~al.}
\newblock Early 2017 observations of {TRAPPIST}-1 with {Spitzer}.
\newblock {\em Mon. Not. R. Astron. Soc.} {\bf 2018},
{\em 475},~3577--3597. [\href{http://dx.doi.org/10.1093/mnras/sty051}{CrossRef}]

\bibitem[Pinamonti et~al.(2018)Pinamonti, Damasso, Marzari, Sozzetti, Desidera,
Maldonado, Scandariato, Affer, Lanza, Bignamini, Bonomo, Borsa, Claudi,
Cosentino, Giacobbe, González-Álvarez, González~Hernández, Gratton, Leto,
Malavolta, Martinez~Fiorenzano, Micela, Molinari, Pagano, Pedani, Perger,
Piotto, Rebolo, Ribas, Suárez~Mascareño, and
Toledo-Padrón]{pinamonti_hades_2018}
Pinamonti, M.; Damasso, M.; Marzari, F.; Sozzetti, A.; Desidera, S.; Maldonado,
J.; Scandariato, G.; Affer, L.; Lanza, A.F.; Bignamini, A.;  et~al.
\newblock The {HADES} {RV} {Programme} with {HARPS}-{N} at {TNG}. {VIII}.
{GJ15A}: A multiple wide planetary system sculpted by binary interaction.
\newblock {\em Astron. Astrophys.} {\bf 2018}, {\em 617},~A104. [\href{http://dx.doi.org/10.1051/0004-6361/201732535}{CrossRef}]

\bibitem[Anglada-Escudé et~al.(2013)Anglada-Escudé, Tuomi, Gerlach, Barnes,
Heller, Jenkins, Wende, Vogt, Butler, Reiners, and
Jones]{anglada-escude_dynamically-packed_2013}
Anglada-Escudé, G.; Tuomi, M.; Gerlach, E.; Barnes, R.; Heller, R.; Jenkins,
J.S.; Wende, S.; Vogt, S.S.; Butler, R.P.; Reiners, A.;  et~al.
\newblock A dynamically-packed planetary system around {GJ} {667C} with three
super-{Earths} in its habitable zone.
\newblock {\em Astron. Astrophys.} {\bf 2013}, {\em 556},~A126. [\href{http://dx.doi.org/10.1051/0004-6361/201321331}{CrossRef}]

\bibitem[Bauer et~al.(2020)Bauer, Zechmeister, Kaminski, Rodríguez~López,
Caballero, Azzaro, Stahl, Kossakowski, Quirrenbach, Becerril~Jarque,
Rodríguez, Amado, Seifert, Reiners, Schäfer, Ribas, Béjar,
Cortés-Contreras, Dreizler, Hatzes, Henning, Jeffers, Kürster, Lafarga,
Montes, Morales, Schmitt, Schweitzer, and Solano]{bauer_carmenes_2020}
Bauer, F.F.; Zechmeister, M.; Kaminski, A.; Rodríguez~López, C.; Caballero,
J.A.; Azzaro, M.; Stahl, O.; Kossakowski, D.; Quirrenbach, A.;
Becerril~Jarque, S.;  et~al.
\newblock The {CARMENES} search for exoplanets around {M} dwarfs. {Measuring}
precise radial velocities in the near infrared: {The} example of the
super-{Earth} {CD} {Cet} b.
\newblock {\em Astron. Astrophys.} {\bf 2020},
{\em 640},~A50. [\href{http://dx.doi.org/10.1051/0004-6361/202038031}{CrossRef}]

\bibitem[Hidalgo et~al.(2020)Hidalgo, Pallé, Alonso, Gandolfi, Fridlund,
Nowak, Luque, Hirano, Justesen, Cochran, Barragán, Spina, Rodler, Albrecht,
Anderson, Amado, Bryant, Caballero, Cabrera, Csizmadia, Dai, De~Leon, Deeg,
Eigmuller, Endl, Erikson, Esposito, Figueira, Georgieva, Grziwa, Guenther,
Hatzes, Hjorth, Hoeijmakers, Kabath, Korth, Kuzuhara, Lafarga, Lampon, Leão,
Livingston, Mathur, Montañes-Rodriguez, Morales, Murgas, Nagel, Narita,
Nielsen, Patzold, Persson, Prieto-Arranz, Quirrenbach, Rauer, Redfield,
Reiners, Ribas, Smith, Šubjak, Van~Eylen, and Wilson]{hidalgo_three_2020}
Hidalgo, D.; Pallé, E.; Alonso, R.; Gandolfi, D.; Fridlund, M.; Nowak, G.;
Luque, R.; Hirano, T.; Justesen, A.B.; \mbox{Cochran, W.D.;  et~al.}
\newblock Three planets transiting the evolved star EPIC 249893012: A hot
8.8-M$_\oplus$ super-Earth and two warm 14.7 and 10.2-M$_\oplus$
sub-Neptunes.
\newblock {\em Astron. Astrophys.} {\bf 2020}, {\em 636},~A89. [\href{http://dx.doi.org/10.1051/0004-6361/201937080}{CrossRef}]

\bibitem[Des~Etangs et~al.(2008)Des~Etangs, Pont, Vidal-Madjar, and
Sing]{des2008rayleigh}
Des~Etangs, A.L.; Pont, F.; Vidal-Madjar, A.; Sing, D.
\newblock Rayleigh scattering in the transit spectrum of HD 189733b.
\newblock {\em Astron. Astrophys.} {\bf 2008}, {\em 481},~L83--L86. [\href{http://dx.doi.org/10.1051/0004-6361:200809388}{CrossRef}]

\bibitem[Stephens et~al.(2020)Stephens, Vannah, and Gleiser]{Stephens2020}
Stephens, M.; Vannah, S.; Gleiser, M.
\newblock Informational approach to cosmological parameter estimation.
\newblock {\em Phys. Rev. D} {\bf 2020}, {\em 102},~{}{123514}. [\href{http://dx.doi.org/10.1103/PhysRevD.102.123514}{CrossRef}]

\bibitem[Thakur et~al.(2020)Thakur, Gleiser, Kumar, and
Gupta]{thakur2020configurational}
Thakur, P.; Gleiser, M.; Kumar, A.; Gupta, R.
\newblock Configurational entropy of optical bright similariton in tapered
graded-index waveguide.
\newblock {\em Phys. Lett. A} {\bf 2020}, {\em 384},~{}{126461}. [\href{http://dx.doi.org/10.1016/j.physleta.2020.126461}{CrossRef}]


\bibitem[Bernardini and da~Rocha(2019)]{bernardini2019}
Bernardini, A.E.; da~Rocha, R.
\newblock Cosmological comoving behavior of the configurational entropy.
\newblock {\em Phys. Lett. B} {\bf 2019}, {\em 796},~{}{107--111}. [\href{http://dx.doi.org/10.1016/j.physletb.2019.07.028}{CrossRef}]


\bibitem[Gleiser and Sowinski(2013)]{Gleiser2013}
Gleiser, M.; Sowinski, D.
\newblock Information-entropic stability bound for compact objects: Application
to Q-balls and the Chandrasekhar limit of polytropes.
\newblock {\em Phys. Lett. B} {\bf 2013}, {\em 727},~272--275. [\href{http://dx.doi.org/10.1016/j.physletb.2013.10.005}{CrossRef}]

\bibitem[Gleiser and Sowinski(2018)]{Gleiser2018}
Gleiser, M.; Sowinski, D.
\newblock Configurational information approach to instantons and false vacuum
decay in $D$-dimensional spacetime.
\newblock {\em Phys. Rev. D} {\bf 2018}, {\em 98},~056026. [\href{http://dx.doi.org/10.1103/PhysRevD.98.056026}{CrossRef}]

\bibitem[Bernardini et~al.(2017)Bernardini, Braga, and
da~Rocha]{bernardini2017}
Bernardini, A.E.; Braga, N.R.F.; da~Rocha, R.
\newblock Configurational entropy of glueball states.
\newblock {\em Phys. Lett. B} {\bf 2017}, {\em 765},~81--85. [\href{http://dx.doi.org/10.1016/j.physletb.2016.12.007}{CrossRef}]

\bibitem[Braga and da~Rocha(2017)]{braga2017}
Braga, N.R.F.; da~Rocha, R.
\newblock Configurational entropy of anti-de Sitter black holes.
\newblock {\em Phys. Lett. B} {\bf 2017}, {\em 767},~386--391. [\href{http://dx.doi.org/10.1016/j.physletb.2017.02.031}{CrossRef}]

\bibitem[Sandford et~al.(2021)Sandford, Kipping, and
Collins]{sandford2021planetary}
Sandford, E.; Kipping, D.; Collins, M.
\newblock On planetary systems as ordered sequences.
\newblock {\em Mon. Not. R. Astron. Soc.} {\bf 2021},
{\em 505},~2224--2246. [\href{http://dx.doi.org/10.1093/mnras/stab1480}{CrossRef}]

\bibitem[Guttenberg et~al.(2021)Guttenberg, Chen, Mochizuki, and
Cleaves]{Guttenberg2021}
Guttenberg, N.; Chen, H.; Mochizuki, T.; Cleaves, H.
\newblock Classification of the Biogenicity of Complex Organic Mixtures for the
Detection of Extraterrestrial Life.
\newblock {\em Life} {\bf 2021}, {\em 11},~234. [\href{http://dx.doi.org/10.3390/life11030234}{CrossRef}]

\bibitem[Marshall et~al.(2021)Marshall, Mathis, Carrick, Keenan, Cooper,
Graham, Craven, Gromski, Moore, Walker, and Cronin]{Marshall2021}
Marshall, S.M.; Mathis, C.; Carrick, E.; Keenan, G.; Cooper, G.J.T.; Graham,
H.; Craven, M.; Gromski, P.S.; Moore, D.G.; \mbox{Walker, S.I.;  et~al.}
\newblock Identifying molecules as biosignatures with assembly theory and mass
spectrometry.
\newblock {\em Nat. Commun.} {\bf 2021}, {\em 12},~3033. [\href{http://dx.doi.org/10.1038/s41467-021-23258-x}{CrossRef}]

\bibitem[Chou et~al.(2021)Chou, Mahaffy, Trainer, Eigenbrode, Arevalo,
Brinckerhoff, Getty, Grefenstette, Poian, Fricke, Kempes, Marlow, Lollar,
Graham, and Johnson]{Chou2021}
Chou, L.; Mahaffy, P.; Trainer, M.; Eigenbrode, J.; Arevalo, R.; Brinckerhoff,
W.; Getty, S.; Grefenstette, N.; Poian, V.D.; \mbox{Fricke, G.M.;  et~al.}
\newblock Planetary Mass Spectrometry for Agnostic Life Detection in the Solar
System.
\newblock {\em Front. Astron. Space Sci.} {\bf 2021}, {\em 8}, 755100. [\href{http://dx.doi.org/10.3389/fspas.2021.755100}{CrossRef}]

\bibitem[Bartlett et~al.(2022)Bartlett, Li, Gu, Sinapayen, Fan, Natraj, Jiang,
Crisp, and Yung]{Bartlett2022}
Bartlett, S.; Li, J.; Gu, L.; Sinapayen, L.; Fan, S.; Natraj, V.; Jiang, J.H.;
Crisp, D.; Yung, Y.L.
\newblock Assessing planetary complexity and potential agnostic biosignatures
using epsilon machines.
\newblock {\em Nat. Astron.} {\bf 2022}, {\em 6},~387--392. [\href{http://dx.doi.org/10.1038/s41550-021-01559-x}{CrossRef}]

\bibitem[Fields et~al.(2023)Fields, Gupta, and Sandora]{Fields_2023}
Fields, B.; Gupta, S.; Sandora, M.
\newblock Information gain as a tool for assessing biosignature missions.
\newblock {\em Int. J. Astrobiol.} {\bf 2023}, {\em
22},~583–607. [\href{http://dx.doi.org/10.1017/S1473550423000150}{CrossRef}]

\bibitem[Sowinski et~al.(2024)Sowinski, Ghoshal, and Frank]{sowinski2024}
Sowinski, D.R.; Ghoshal, G.; Frank, A.
\newblock Exo-Daisy World: Revisiting Gaia Theory through an Informational
Architecture Perspective. \emph{arXiv}  \textbf{2024}, {}{arXiv:2411.03421}.
 [\href{http://dx.doi.org/10.48550/arXiv.2411.03421}{CrossRef}]

\bibitem[Wong et~al.(2023)Wong, Prabhu, Williams, Morrison, and
Hazen]{Wong2023}
Wong, M.L.; Prabhu, A.; Williams, J.; Morrison, S.M.; Hazen, R.M.
\newblock Toward Network‐Based Planetary Biosignatures: Atmospheric Chemistry
as Unipartite, Unweighted, Undirected Networks.
\newblock {\em J. Geophys. Res. Planets} {\bf 2023}, {\em 128}, e2022JE007658. [\href{http://dx.doi.org/10.1029/2022JE007658}{CrossRef}]

\bibitem[Fisher et~al.(2023)Fisher, Janin, and
Walker]{fisher2023complexsystemsapproachexoplanet}
Fisher, T.; Janin, E.; Walker, S.I.
\newblock A Complex Systems Approach to Exoplanet Atmospheric Chemistry: New
Prospects for Ruling Out the Possibility of Alien Life-As-We-Know-It. \emph{arXiv} \textbf{2023}, arXiv:2310.05359. [\href{http://dx.doi.org/10.48550/arXiv.2310.05359}{CrossRef}]

\bibitem[Guez and Claire(2024)]{guez2024}
Guez, I.A.; Claire, M.
\newblock Reading Between the Rainbows: Comparative Exoplanet Characterisation
through Molecule Agnostic Spectral Clustering. \emph{arXiv} {\bf 2024}, arXiv:2410.16986. [\href{http://dx.doi.org/10.48550/arXiv.2410.16986}{CrossRef}]

\bibitem[Lovelock(1965)]{Lovelock1965}
Lovelock, J.E.
\newblock A physical basis for life detection experiments.
\newblock {\em Nature} {\bf 1965}, {\em 207},~568--570. [\href{http://dx.doi.org/10.1038/207568a0}{CrossRef}]

\bibitem[Lederberg(1965)]{Lederberg1965}
Lederberg, J.
\newblock Signs of life.
\newblock {\em Nature} {\bf 1965}, {\em 207},~9--13. [\href{http://dx.doi.org/10.1038/207009a0}{CrossRef}]

\bibitem[Cockell et~al.(2009)Cockell, L\'{e}ger, Fridlund, Herbst, Kaltenegger,
Absil, Beichman, Benz, Blanc, Brack, Chelli, Colangeli, Cottin, Foresto,
Danchi, Defr\`{e}re, Herder, Eiroa, Greaves, Henning, Johnston, Jones,
Labadie, Lammer, Launhardt, Lawson, Lay, LeDuigou, Liseau, Malbet, Martin,
Mawet, Mourard, Moutou, Mugnier, Ollivier, Paresce, Quirrenbach, Rabbia,
Raven, Rottgering, Rouan, Santos, Selsis, Serabyn, Shibai, Tamura,
Thi\'{e}baut, Westall, and White]{cockell2009}
Cockell, C.; L\'{e}ger, A.; Fridlund, M.; Herbst, T.; Kaltenegger, L.; Absil,
O.; Beichman, C.; Benz, W.; Blanc, M.; Brack, A.;  et~al.
\newblock Darwin—A Mission to Detect and Search for Life on Extrasolar
Planets.
\newblock {\em Astrobiology} {\bf 2009}, {\em 9},~{}{1--22}
. [\href{http://dx.doi.org/10.1089/ast.2007.0227}{CrossRef}] [\href{http://www.ncbi.nlm.nih.gov/pubmed/19203238}{PubMed}]

\bibitem[Krissansen-Totton et~al.(2018)Krissansen-Totton, Olson, and
Catling]{krissansentotton2018}
Krissansen-Totton, J.; Olson, S.; Catling, D.C.
\newblock Disequilibrium biosignatures over Earth history and implications for
detecting exoplanet life.
\newblock {\em Sci. Adv.} {\bf 2018}, {\em 4},~eaao5747. [\href{http://dx.doi.org/10.1126/sciadv.aao5747}{CrossRef}]

\bibitem[Young et~al.(2024)Young, Robinson, Krissansen-Totton, Schwieterman,
Wogan, Way, Sohl, Arney, Reinhard, Line, Catling, and Windsor]{young2024}
Young, A.V.; Robinson, T.D.; Krissansen-Totton, J.; Schwieterman, E.W.; Wogan,
N.F.; Way, M.J.; Sohl, L.E.; Arney, G.N.; Reinhard, C.T.; Line, M.R.;  et~al.
\newblock Inferring chemical disequilibrium biosignatures for Proterozoic
Earth-like exoplanets.
\newblock {\em Nat. Astron.} {\bf 2024}, {\em 8},~{}{101--110}. [\href{http://dx.doi.org/10.1038/s41550-023-02145-z}{CrossRef}]


\bibitem[Herbst et~al.(2024)Herbst, Bartenschlager, Grenfell, Iro, Sinnhuber,
Taysum, Wunderlich, Engelbrecht, Light, Moloto, Harre, Rauer, and
Schreier]{Herbst_2024}
Herbst, K.; Bartenschlager, A.; Grenfell, J.L.; Iro, N.; Sinnhuber, M.; Taysum,
B.; Wunderlich, F.; Engelbrecht, N.E.; Light, J.; Moloto, K.D.;  et~al.
\newblock Impact of Cosmic Rays on Atmospheric Ion Chemistry and Spectral
Transmission Features of TRAPPIST-1e.
\newblock {\em  Astrophys. J.} {\bf 2024}, {\em 961},~164. [\href{http://dx.doi.org/10.3847/1538-4357/ad0895}{CrossRef}]

\bibitem[Doyon(2024)]{doyon2024temperaterockyplanetsm}
Doyon, R.
\newblock Do Temperate Rocky Planets Around M Dwarfs have an Atmosphere?
\newblock \emph{arXiv}  \textbf{2024}, arXiv:2403.12617v3. [\href{http://dx.doi.org/10.48550/arXiv.2403.12617}{CrossRef}]

\bibitem[Lin and Kaltenegger(2022)]{lin2022}
Lin, Z.; Kaltenegger, L.
\newblock High-resolution spectral models of TRAPPIST-1e seen as a Pale Blue
Dot for ELT and JWST observations.
\newblock {\em Mon. Not. R. Astron. Soc.} {\bf 2022},
{\em 516},~3167--3174. [\href{http://dx.doi.org/10.1093/mnras/stac2536}{CrossRef}]

\bibitem[Fauchez et~al.(2019)Fauchez, Turbet, Villanueva, Wolf, Arney,
Kopparapu, Lincowski, Mandell, Wit, Pidhorodetska, Domagal-Goldman, and
Stevenson]{Fauchez_2019}
Fauchez, T.J.; Turbet, M.; Villanueva, G.L.; Wolf, E.T.; Arney, G.; Kopparapu,
R.K.; Lincowski, A.; Mandell, A.; Wit, J.d.; Pidhorodetska, D.;  et~al.
\newblock Impact of Clouds and Hazes on the Simulated JWST Transmission Spectra
of Habitable Zone Planets in the TRAPPIST-1 System.
\newblock {\em  Astrophys. J.} {\bf 2019}, {\em 887},~194. [\href{http://dx.doi.org/10.3847/1538-4357/ab5862}{CrossRef}]

\bibitem[Sarkar and Madhusudhan(2021)]{sarkar_jexosim_2021}
Sarkar, S.; Madhusudhan, N.
\newblock {JexoSim} 2.0: {End}-to-{End} {JWST} {Simulator} for {Exoplanet}
{Spectroscopy}---{Implementation} and {Case} {Studies}.
\newblock {\em Mon. Not. R. Astron. Soc.} {\bf 2021},
{\em 508},~433--452. [\href{http://dx.doi.org/10.1093/mnras/stab2472}{CrossRef}]

\bibitem[{Batalha} et~al.(2017){Batalha}, {Mandell}, {Pontoppidan},
{Stevenson}, {Lewis}, {Kalirai}, {Earl}, {Greene}, {Albert}, and
{Nielsen}]{batalha2017}
{Batalha}, N.E.; {Mandell}, A.; {Pontoppidan}, K.; {Stevenson}, K.B.; {Lewis},
N.K.; {Kalirai}, J.; {Earl}, N.; {Greene}, T.; {Albert}, L.; \mbox{{Nielsen}, L.D.}
\newblock {PandExo: A Community Tool for Transiting Exoplanet Science with JWST
\& HST}.
\newblock {\em Publ. Astron. Soc. Pac.} {\bf 2017}, {\em 129},~064501. [\href{http://dx.doi.org/10.1088/1538-3873/aa65b0}{CrossRef}]

\bibitem[Ahrer et~al.(2022)Ahrer, Alderson, Batalha, Batalha, Bean, Beatty,
Bell, Benneke, Berta-Thompson, Carter, Crossfield, Espinoza, Feinstein,
Fortney, Gibson, Goyal, Kempton, Kirk, Kreidberg, L{\'o}pez-Morales, Line,
Lothringer, Moran, Mukherjee, Ohno, Parmentier, Piaulet, Rustamkulov,
Schlawin, Sing, Stevenson, Wakeford, Allen, Birkmann, Brande, Crouzet,
Cubillos, Damiano, D{\'e}sert, Gao, Harrington, Hu, Kendrew, Knutson, Lagage,
Leconte, Lendl, MacDonald, May, Miguel, Molaverdikhani, Moses, Murray,
Nehring, Nikolov, Petit dit de~la Roche, Radica, Roy, Stassun, Taylor,
Waalkes, Wachiraphan, Welbanks, Wheatley, Aggarwal, Alam, Banerjee, Barstow,
Blecic, Casewell, Changeat, Chubb, Col{\'o}n, Coulombe, Daylan, de~Val-Borro,
Decin, Dos~Santos, Flagg, France, Fu, Mu{\~n}oz, Gizis, Glidden, Grant, Heng,
Henning, Hong, Inglis, Iro, Kataria, Komacek, Krick, Lee, Lewis, Lillo-Box,
Lustig-Yaeger, Mancini, Mandell, Mansfield, Marley, Mikal-Evans, Morello,
Nixon, Ceballos, Piette, Powell, Rackham, Ramos-Rosado, Rauscher, Redfield,
Rogers, Roman, Roudier, Scarsdale, Shkolnik, Southworth, Spake, Steinrueck,
Tan, Teske, Tremblin, Tsai, Tucker, Turner, Valenti, Venot, Waldmann,
Wallack, Zhang, Zieba, and Team]{aher2022}
Ahrer, E.M.; Alderson, L.; Batalha, N.M.; Batalha, N.E.; Bean, J.L.; Beatty,
T.G.; Bell, T.J.; Benneke, B.; Berta-Thompson, Z.K.; Carter, A.L.;  et~al.
\newblock Identification of carbon dioxide in an exoplanet atmosphere.
\newblock {\em Nature} {\bf {}{2023}
}, \emph{614}, 649--652. [\href{http://dx.doi.org/10.1038/s41586-022-05269-w}{CrossRef}]

\bibitem[{Bell} et~al.(2023){Bell}, {Welbanks}, {Schlawin}, {Line}, {Fortney},
{Greene}, {Ohno}, {Parmentier}, {Rauscher}, {Beatty}, {Mukherjee}, {Wiser},
{Boyer}, {Rieke}, and {Stansberry}]{bell2023}
{Bell}, T.J.; {Welbanks}, L.; {Schlawin}, E.; {Line}, M.R.; {Fortney}, J.J.;
{Greene}, T.P.; {Ohno}, K.; {Parmentier}, V.; {Rauscher}, E.; {Beatty}, T.G.;
et~al.
\newblock {Methane throughout the atmosphere of the warm exoplanet WASP-80b}.
\newblock {\em Nature} {\bf 2023}, {\em 623},~709--712. [\href{http://dx.doi.org/10.1038/s41586-023-06687-0}{CrossRef}]

\bibitem[{Madhusudhan} et~al.(2023){Madhusudhan}, {Sarkar}, {Constantinou},
{Holmberg}, {Piette}, and {Moses}]{madhusudhan2023}
{Madhusudhan}, N.; {Sarkar}, S.; {Constantinou}, S.; {Holmberg}, M.; {Piette},
A.A.A.; {Moses}, J.I.
\newblock {Carbon-bearing Molecules in a Possible Hycean Atmosphere}.
\newblock {\em Astrophys. J.} {\bf 2023}, {\em 956},~L13. [\href{http://dx.doi.org/10.3847/2041-8213/acf577}{CrossRef}]

\bibitem[Schmidt et~al.(2025)Schmidt, MacDonald, Tsai, Radica, Wang, Ahrer,
Bell, Fisher, Thorngren, Wogan, May, Ferrari, Bennett, Rustamkulov,
López-Morales, and Sing]{schmidt2025}
Schmidt, S.P.; MacDonald, R.J.; Tsai, S.M.; Radica, M.; Wang, L.C.; Ahrer,
E.M.; Bell, T.J.; Fisher, C.; Thorngren, D.P.; \mbox{Wogan, N.;  et~al.}
\newblock A Comprehensive Reanalysis of K2-18 b's JWST NIRISS+NIRSpec
Transmission Spectrum. \emph{arXiv}  \textbf{2025}, arXiv:2501.18477. [\href{http://dx.doi.org/10.48550/arXiv.2501.18477}{CrossRef}]

\bibitem[Vannah(2022)]{vannah-phd}
Vannah, S.
\newblock Information Entropic Content of Astrophysical Spectra: Applications
to Cosmology and Astrobiology.
\newblock Ph.D. Thesis, Dartmouth College, {}{Hanover, NH, USA}
, 2022.

\bibitem[{Beichman} and {Greene}(2018)]{beichman2018}
{Beichman}, C.A.; {Greene}, T.P.
\newblock {A White Paper Submitted to The National Academy of Science's
Committee on Exoplanet Science Strategy: Observing Exoplanets with the James
Webb Space Telescope}.
\newblock {\em arXiv} {\bf 2018}, {}{arXiv:1803.03730}.
\newblock{\url{
.
}} [\href{http://dx.doi.org/10.48550/arXiv.1803.03730}{CrossRef}]


\end{thebibliography}
\end{document}